\documentclass[review]{elsarticle}

\usepackage{graphicx}
\usepackage{amssymb}
\usepackage{amsfonts}
\usepackage{amsmath}
\usepackage{amsbsy}
\usepackage{natbib}
\usepackage{comment}
\usepackage{color}
\usepackage{float}

\renewcommand{\vec}[1]{\mbox{\boldmath$\mathrm{#1}$}}
\newcommand{\be}{\begin{equation}}
\newcommand{\ee}{\end{equation}}
\newcommand{\ben}{\begin{eqnarray}}
\newcommand{\een}{\end{eqnarray}}

\begin{document}

\begin{frontmatter}

\title{Electric control of emergent magnonic spin current and dynamic multiferroicity in magnetic insulators at finite temperatures}

\author{Xi-guang Wang$^{1,2}$, L. Chotorlishvili$^{1}$,  Guang-hua Guo$^2$, J. Berakdar$^1$}

\address{$^1$Institut f\"ur Physik, Martin-Luther Universit\"at Halle-Wittenberg, 06099 Halle/Saale, Germany \\
	$^2$School of Physics and Electronics, Central South University, Changsha 410083, China}

\begin{abstract}
Conversion of thermal energy into magnonic spin currents and/or effective electric polarization promises new device functionalities. A versatile approach is presented here for generating and controlling open circuit magnonic spin currents and an effective multiferroicity at a uniform temperature with the aid of spatially inhomogeneous, external, static electric fields. This field applied to a ferromagnetic insulator  with a Dzyaloshinskii-Moriya type coupling changes locally the  magnon  dispersion and modifies  the density of  thermally excited magnons in a region of the scale of the field inhomogeneity. The resulting gradient in the magnon density can be viewed as a gradient in the effective magnon  temperature. This effective thermal gradient together with local magnon dispersion result  in an open-circuit, electric field controlled  magnonic spin current. In fact, for  a moderate variation in the external electric field the predicted  magnonic spin current is on the scale of the spin (Seebeck) current generated by a comparable external  temperature gradient.
Analytical methods supported by full-fledge numerics  confirm that both, a finite temperature and an inhomogeneous electric field are necessary for this emergent non-equilibrium  phenomena.
The proposal can be integrated  in  magnonic and multiferroic circuits, for instance to convert heat into electrically controlled pure spin current using for example  nanopatterning,  without the need to generate large thermal gradients on the nanoscale.
\end{abstract}


\end{frontmatter}

\section{Introduction}

\begin{figure}
	\centering
	\includegraphics[width=0.38\textwidth]{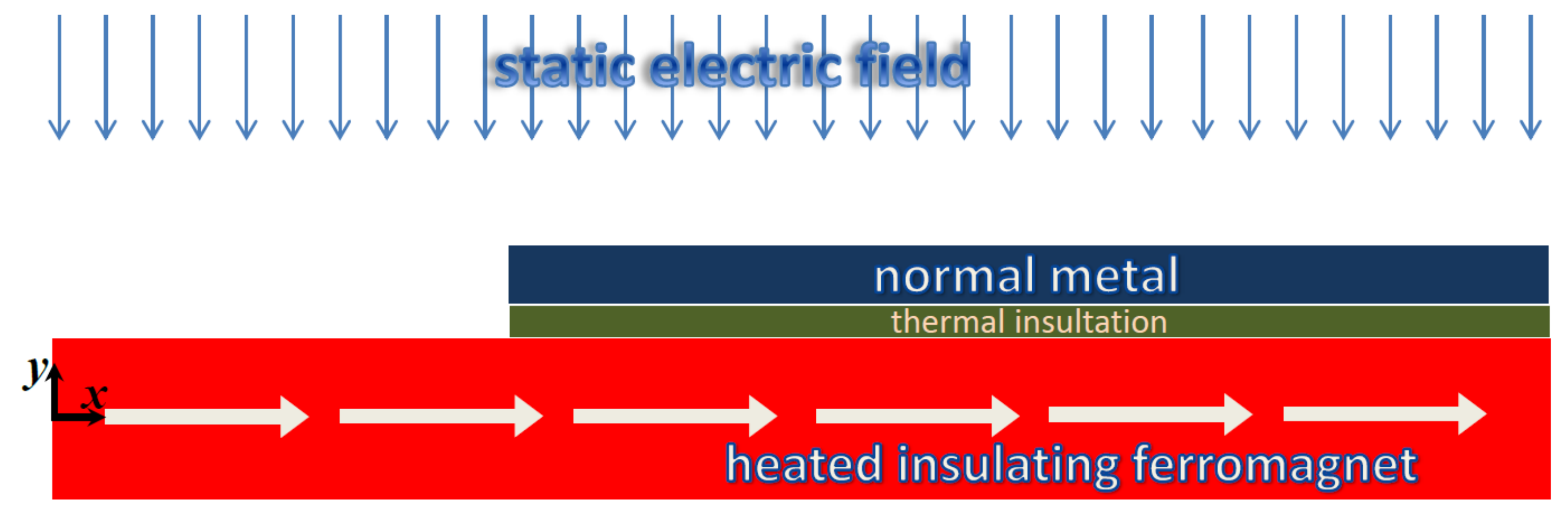}
	\vspace*{0.3cm}\\
	\includegraphics[width=0.43\textwidth]{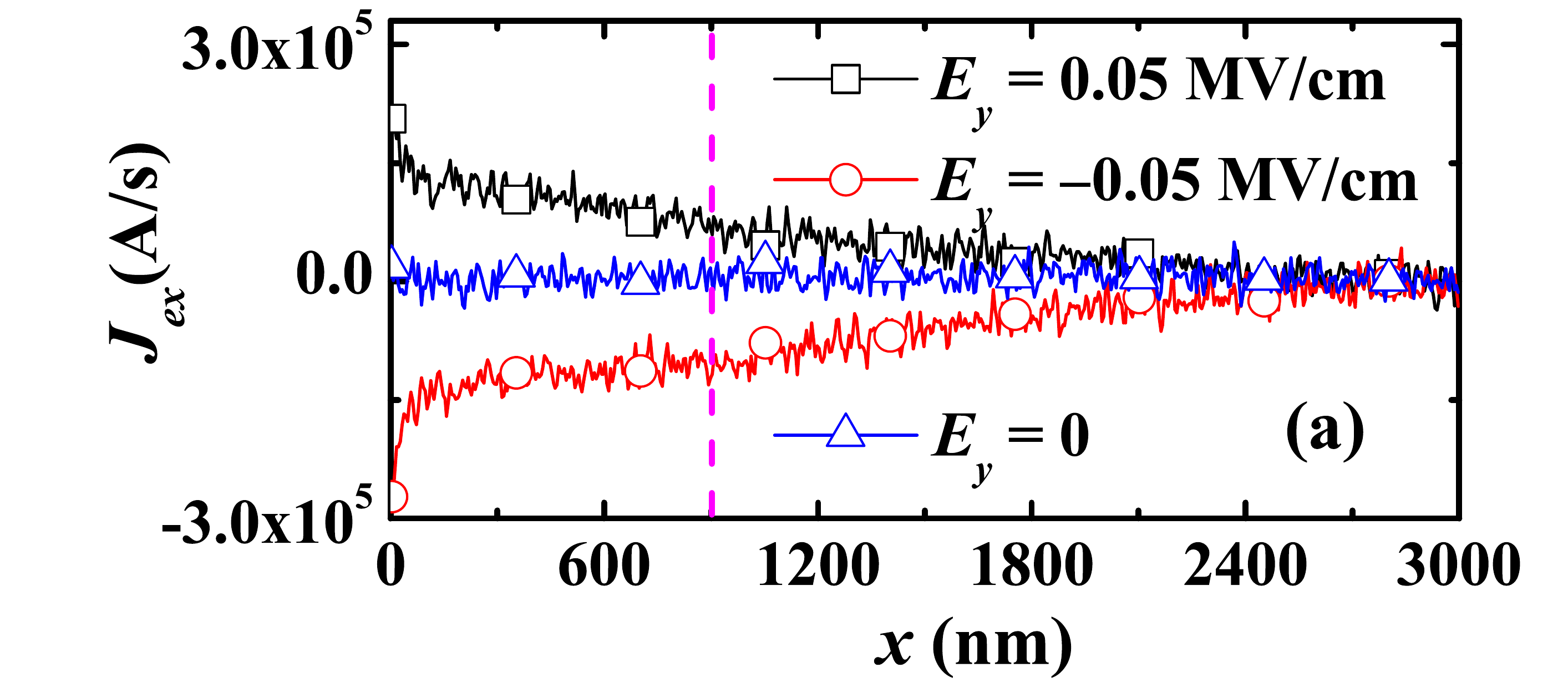}
	\includegraphics[width=0.46\textwidth]{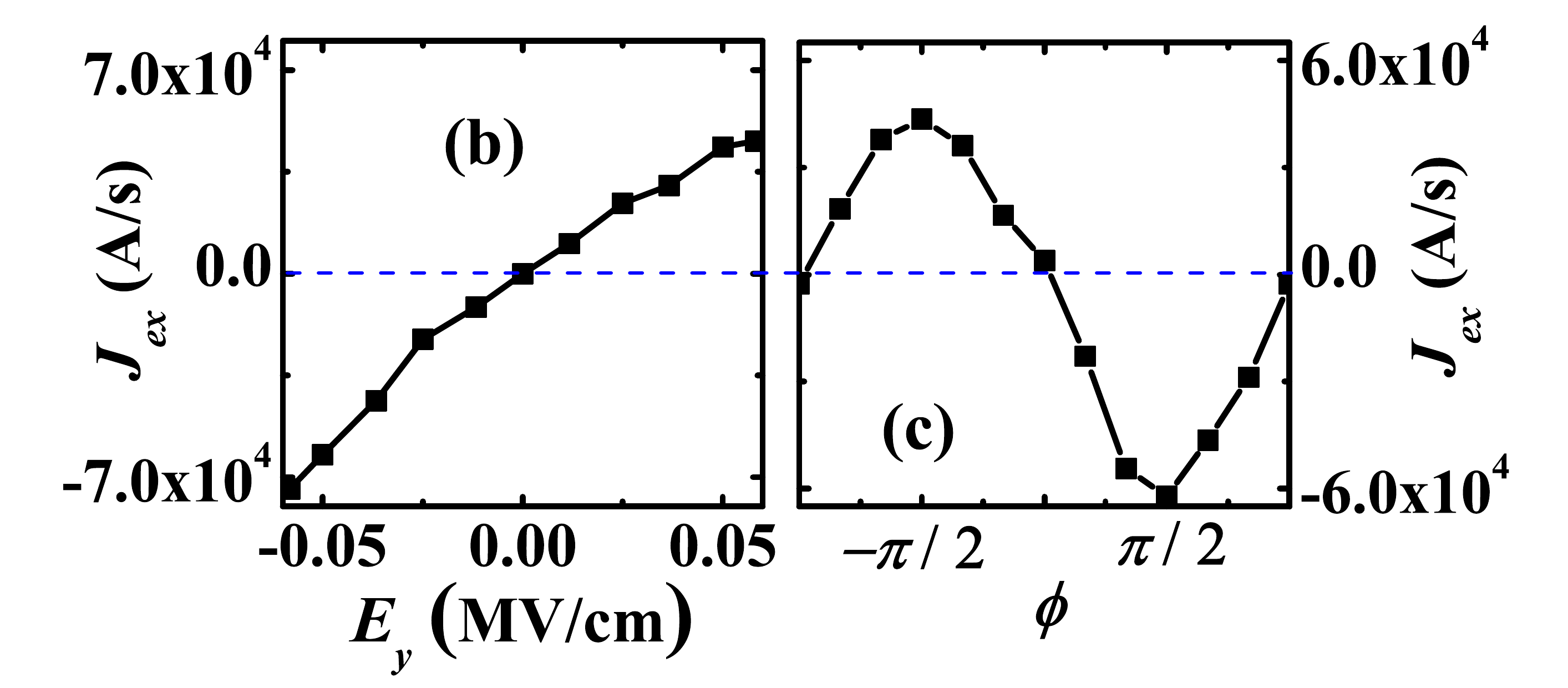}
	\caption{Top panel shows  schematically   the studied structure consisting of an insulating ferromagnet heated uniformly to the temperature $T$ and subjected to a static electric field $ E_{y}$ which is shielded in parts of the ferromagnet, for instance by a normal metallic cap layer.
		(a) $ J_{ex} $,  the exchange spin current for $ E_{y}=0 $  (blue triangles), $ E_{y}=0.05 $ MV/cm (black squares), and $ E_{y}=-0.05 $ MV/cm (red circles).  $ E_{y}$ acts on
		$ 0 \leq x \leq 900 $ nm.  \textbf{ Symbols are to guide the eye and distinguish the curves.} (b) When the local magnetization is parallel to the $ z $ axis ($ \theta = \pi / 2 $ and $ \phi = \pi / 2 $), averaged exchange spin current $ J_{ex} $ in the region $ 1000 \text{ nm} \leq x \leq 2000 \text{ nm} $ as a function of $ E_{y} $. (c) When $ E_{y}=-0.05 $ MV/cm and $ \theta = \pi / 2 $, averaged exchange spin current $ J_{ex} $ in the region of $ 1000 \text{ nm} \leq x \leq 2000 \text{ nm} $ as a function of  $\phi $. }
	\label{fig_1}
\end{figure}

A wealth of fascinating phenomena in electronic systems emerge from the
interplay of electronic correlation, symmetry, and the coupled spin-orbital dynamics. A shining example is the emergence of a spin-driven ferroelectricity coupled to a helical magnetic ordering in certain  oxides \cite{KNB,mostovoy,Sergienko,tokurarpp,Tokura,Tokura01,Tokura02,Khomskii,Khomskii01,Khomskii02,Khomskii03,Mats01,JinJ01}.
What about entropy. Can thermal magnetic fluctuations help render  an emergent effective  dynamic multiferroicity? This question addressed here is not only fundamentally important,  it also bears a high potential for electrically-controlled caloritronics, a field which is undergoing a dynamic development
\cite{Uchida1,KaHa10,JiUp13,ScRi14,EtCh14,Schreier,khomeriki,Sukhov,Sukhov01,Barnas,Cornelissen,Jakob,Cornelissen2,VBass11,SRBo12,JBar12,JBar1201}.
In a typical caloritronic setup a magnetically ordered insulator is subjected to a thermal gradient which may generate a pure spin current \cite{Uchida1,KaHa10,JiUp13,ScRi14,EtCh14,Schreier,khomeriki,Sukhov,Sukhov01,Barnas,Cornelissen,Jakob,Cornelissen2,VBass11,SRBo12,JBar12,JBar1201}. At the nanoscale, a clear obstacle is the realization and control of temperature gradients.  Here we work simply at a finite uniform temperature. In addition we apply a spatially inhomogeneous static electric (E) field. Feasible  nanostructuring can be utilized to achieve the E-field inhomogeneity (cf. Fig.  \ref{fig_1}). Also the E-field gradient may be intrinsically present, for example at the interfaces of different materials, involving for instance a ferroelectric cap layer \cite{jia}. Under the conditions identified here,  a steady-state, open circuit magnonic spin current is generated. The underlying mechanism relies on the influence of the electric field gradient on  the spectral characteristics of the spin waves which amounts to
an effective Dzyaloshinskii Moriya (DM) type coupling. Analytical and full numerical calculations confirm and explain this phenomena. The generated spin current at moderate E\-field is on the scale of the spin current which is generated by a thermal gradient within the same temperature range. \\

The paper is organized as follows: In  section \ref{part2} we introduce the model and specify the term emergent effective electric polarization, and how the coupling between the effective electric polarization to the external electric field
gives rise to a dynamical DM interaction. In section \ref{part3}, the energy balance of the ferroelectric subsystem is discussed. In the section \ref{part4} we present  the results and analyze  of the numerical calculations;  section \ref{part5}  discusses  the magnetic dynamics in the region of inhomogeneity of the external electric field   which constitutes  the key issue for the  mechanism of generating  the spin current. These regions are called the interfacial area. In  section \ref{part6} we consider the different spatial orientation of the electric field. The electric generation of spin currents  is compared to the temperature-gradient-induced spin  Seebeck current in  section \ref{part7}. The paper ends with  conclusions and an appendix with technical details.

\section{Theoretical formulation and enhanced dynamical DM}
\label{part2}

In the considered energy range and for the setting described above the relevant dynamic is associated with transversal spin excitations and can be adequately modelled by the dynamics of a magnetic order parameter (denoted by the unit vector field $ \vec{m} $) in a Landau-Ginzburg approach. The total Landau free energy density  $ F_{\text{total}} $  accounts for
the exchange interaction $ A_{ex} (\nabla \vec{m})^2 $, the Zeeman energy $ -\vec{H}_{\text{ext}} \cdot \vec{M} $, and possible magnetic anisotropies. Here $\vec{M} =  M_{s}\vec{m} $ and $M_{s}$ is the saturation magnetization.
Increasing the temperature $T$  activates the magnonic excitations accompanied by random non collinearities
with an associated random emergent effective electric polarization $\vec {\tilde P}$ that averages to zero.
This random effective polarization is named so as it can be stabilized by a moderate static electric field $\vec{E}$ and attains a finite value \cite{Tliu20,VRis}
\begin{equation}
\displaystyle \vec{P} \equiv \langle \vec{\tilde P} \rangle = \langle  c_{E} [(\vec{m} \cdot \nabla) \vec{m} - \vec{m} (\nabla \cdot \vec{m}) ]\rangle .
\label{polarization}
\end{equation}
Here $\langle \dots\rangle$ stands for an ensemble average, and  $c_{E}$ is a weak residual DM coupling constant.
The effective field $\vec{H}_{\text{eff}}$ acting on $\vec{m}$ contains in addition to $-1/(\mu_{0} M_{s}) \delta F_{\text{total}} [\vec{m}]/\delta \vec{m} $ a stochastic contribution due to the external thermal (white) noise.
The term  associated with the electric energy contribution $ H_{D}=-\vec{E} \cdot \vec{\tilde P} $ is also stochastic due to the  random nature of  $\vec{m} $ and  $\nabla \vec{m} $ (and hence of the polarization).
Thermal-averaged quantities are obtained by propagating the stochastic Landau-Lifshitz-Gilbert (LLG) equation \cite{JLGar22} ($ \gamma $ is the gyromagnetic ratio and $ \alpha $ is Gilbert damping which accounts for the relaxation of  spin excitations due to coupling to other degrees of freedom of the system such as the lattice)
\begin{equation}
\displaystyle \frac{\partial \vec{M}}{\partial t} = - \gamma \vec{M} \times (\vec{H}_{\text{eff}} +\vec{h}_{l} - \frac{1}{\mu_0 M_s } \frac{\delta H_D}{\delta \vec{m}})+ \frac{\alpha}{M_{s}} \vec{M} \times \frac{\partial \vec{M}}{\partial t}.
\label{LLG1}
\end{equation}
The predicted effect is of a general nature and its sizable magnitude is  demonstrated here for  iron garnet system
\cite{VRis,ASLo} $ (\mathrm{BiR})_3(\mathrm{FeGa})_5\mathrm{O}_{12} $ (R = Lu, Tm).
With the external magnetic field $ \vec{H}_{\text{ext}} = H_{0} (\text{cos}\theta,\text{sin}\theta\text{cos}\phi,\text{sin}\theta\text{sin}\phi) $
a uniform magnetization  $ \vec{M}_{0} = M_{s} (\text{cos}\theta,\text{sin}\theta\text{cos}\phi,\text{sin}\theta\text{sin}\phi) $
sets in ($ \theta = \pi / 2 $, and $ \phi $ are the polar  and the azimuthal angles).  $T$ is finite.
For  a static electric field $\vec{E} = (0, E_{y}, 0)$, a finite term $\langle -\vec{E} \cdot \vec{\tilde P} \rangle$ mimics effectively  an emergent  $z$ component of a  (non-equilibrium) DM interaction with a strength $D_{E}$  that can be retrieved from the linear response of the  spin current to the electric field.

\begin{figure}
	\centering
	\includegraphics[width=0.49\textwidth]{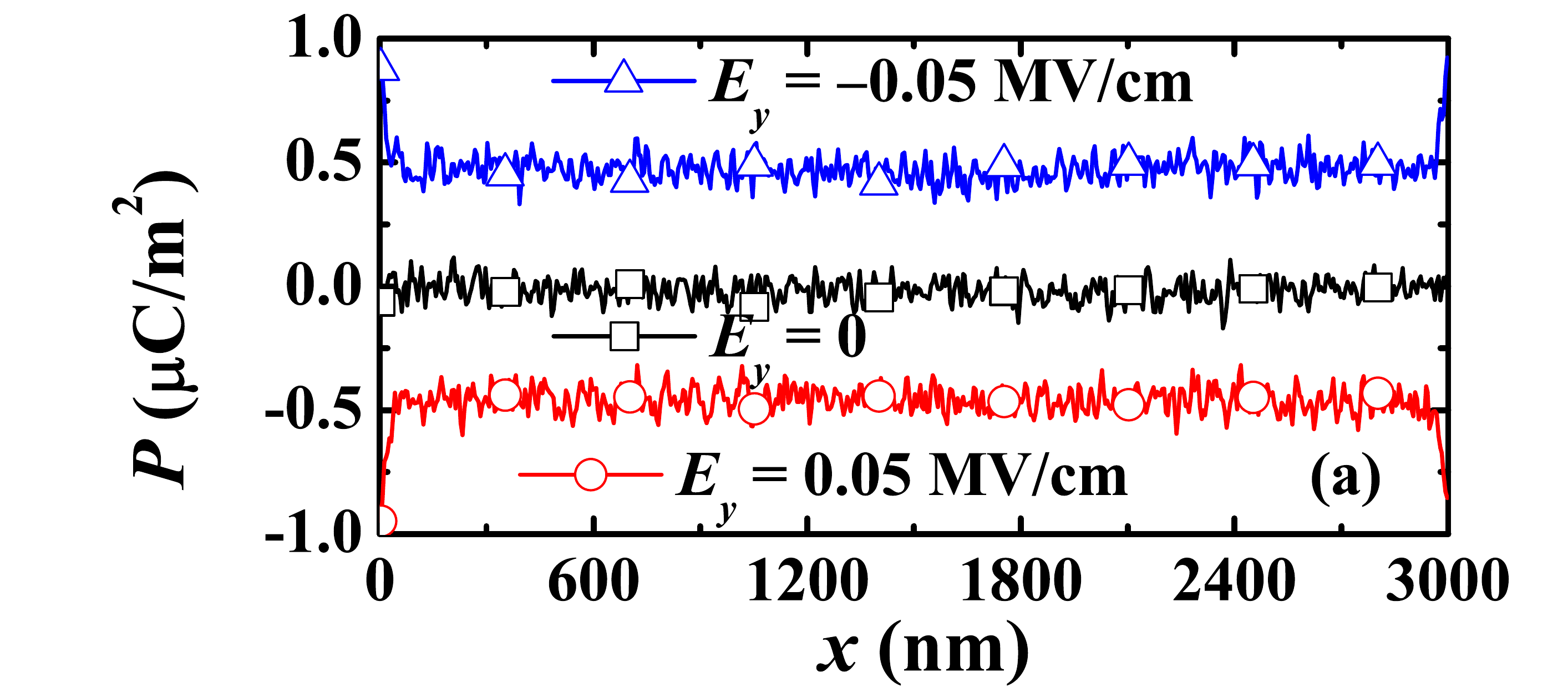}
	\includegraphics[width=0.49\textwidth]{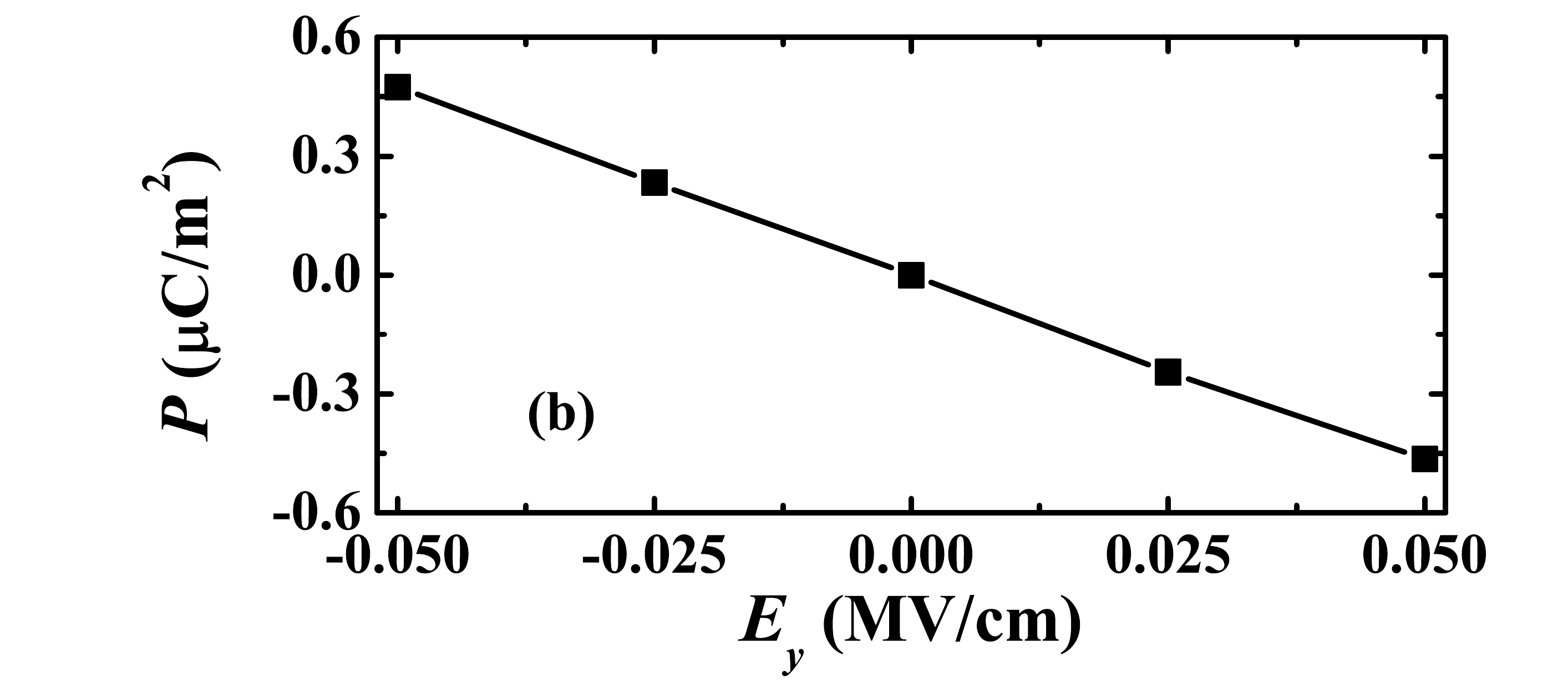}
	\includegraphics[width=0.49\textwidth]{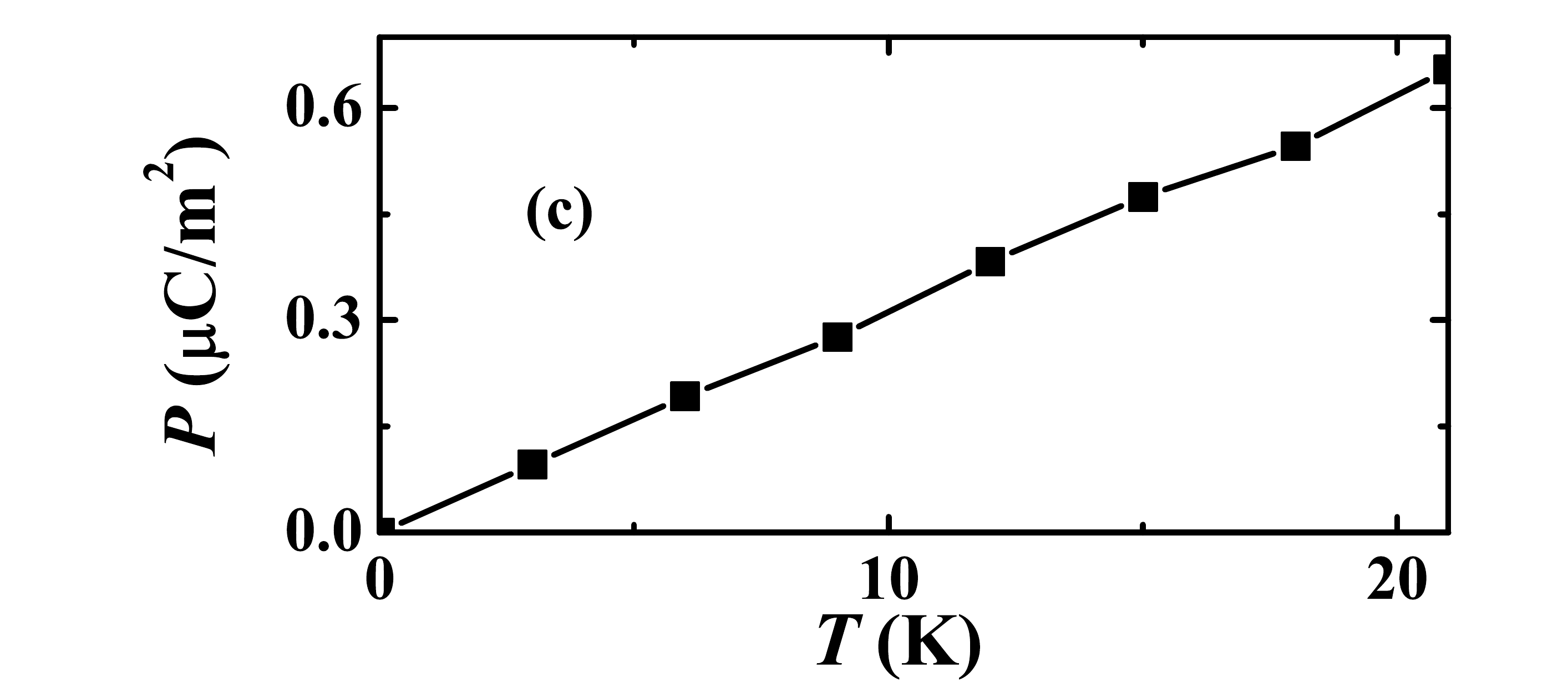}	
	\caption{The local magnetization is parallel to the $ z $ axis ($ \theta = \pi / 2 $ and $ \phi = \pi / 2 $). (a) The value of the effective electric polarization defined by Eq. (\ref{eq_S3}).  The electrical field $ E_{y}=0 $  (black squares), $ E_{y}=0.05 $ MV/cm (red circles) and $ E_{y}=-0.05 $ MV/cm (blue triangles) is applied to the whole system, and the uniform temperature $ T $ is 15 K. (b) The averaged value of the polarization as a function of the electrical field $ E_{y} $ for $ T = 15 $ K. (c) The averaged value of the polarization as a function of $ T $ for $ E_{y}=-0.05 $ MV/cm.}
	\label{fig_S5}
\end{figure}

Let us consider  some results for the emerged net polarization quantified as
$\vec{P} \equiv \langle \vec{\tilde P} \rangle = \langle  c_{E} [(\vec{m} \cdot \nabla) \vec{m} - \vec{m} (\nabla \cdot \vec{m}) ]\rangle$,
\begin{equation}
\displaystyle \vec{P} = -\frac{c_E}{M_{s}^{2}} \langle M_{\theta} \partial_{x} M_{\phi} - M_{\phi} \partial_{x} M_{\theta} \rangle \vec{e}_{\phi}.
\label{eq_S3}
\end{equation}
$ M_{\theta} $ and $ M_{\phi} $ are the thermally activated transversal magnetization components in spherical coordinates, and $ \vec{e}_{\phi} = (0,-\mathrm{sin}\phi, \mathrm{cos}\phi) $. As we see from Eq. (\ref{eq_S3}), the generated polarization is parallel to $ \vec{e}_{\phi} $ and perpendicular to the direction $ \vec{e}_{r} $ of the equilibrium magnetization $ \vec{M}_0 $.
Fig. \ref{fig_S5} demonstrates that the direction of the polarization $ \vec{P} $ is antiparallel to the electric field $ \vec{E} $. When the electric field is directed along the $ +y $ direction ($ E_y > 0 $), the value of $ \vec{P} $ is negative. The polarization reverses sign when $ E_y < 0 $. The polarization vanishes, $ \vec{P}=0$ in absence of the electric field. Micromagnetic simulations confirm a zero polarization when the electrical field is
aligned orthogonal to the $ \vec{e}_{\phi} $ direction.

A stronger amplitude of the electrical field enhances  the polarization $ \vec{P} $, as inferred from  Fig. \ref{fig_S5}(b). Increasing the uniform temperature  enlarges  the magnon density and the polarization $ \vec{P} $, as is shown in Fig. \ref{fig_S5}(c) and Fig. \ref{fig_3}; while  $\lim_{T\to 0},  \vec{P} \to 0$. \\

In the presence of  $E$ field gradient the spin current is driven by \emph{both} the exchange interactions and stochastic DM-type coupling.
The exchange magnonic spin current $ J_{ex} $ is quantified as \cite{KaHa10}
\begin{equation}
\displaystyle J_{ex} = \frac{2 \gamma A_{ex}}{\mu_{0} M_{s}^{2}} \langle M_{\theta} \partial_{x} M_{\phi} - M_{\phi} \partial_{x} M_{\theta} \rangle.
\label{eq_5}
\end{equation}
$ M_{\theta} $ and $ M_{\phi} $ are the thermally activated transversal magnetization angular components.
The magnon spin polarization is   opposite to the local magnetization
and hence  positive spin current $ J_{ex}>0$ is formed by  exchange magnons  propagating along $-x$ direction.\\

The emergent effective DM type term $H_{D} =-\vec{P}\vec{E}$ implies
a  contribution in the spin current of the chiral form $ \partial_{x} J_{D} = \langle \gamma \vec{M} \times \vec{h}_{D}^{eff} \rangle $.
Here $\vec{h}_{D}^{eff}=-1 / (\mu_0 M_s) \delta H_D / \delta \vec{m} $ is the effective field related to the DM term.
When this DM term is small, the local equilibrium magnetization is mostly parallel to the external field
while the effective polarization of the chiral spin current is opposite to the local magnetization. The DM spin current reads \cite{Barnas}
\begin{equation}
\displaystyle J_{D}=-\frac{l_{D} \mathrm{sin} \theta \mathrm{sin} \phi}{2} \langle M_{\theta}^2 + M_{\phi}^2 \rangle.
\label{chiralcurrent}
\end{equation}
Here $ l_{D} = 2 \gamma c_{E} E_{y} /(\mu_{0} M_{s}^{2}) $.

\section{Energy balance of spin current and non-equilibrium magnon density}
\label{part3}

\begin{figure}
	\centering
	\includegraphics[width=0.49\textwidth]{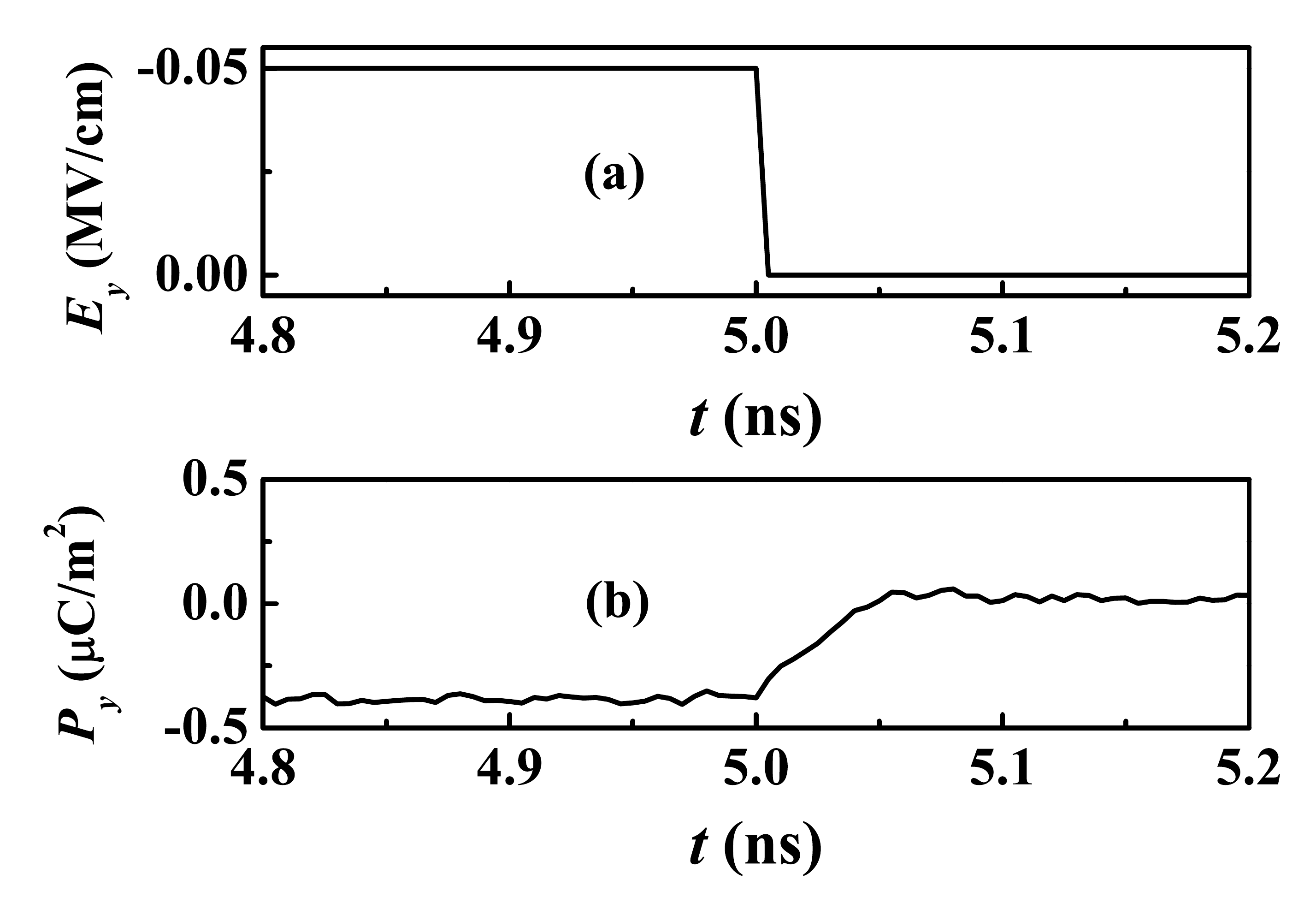}
	\caption{For the electric field switched off at $ t = 5 $ ns (a), the time-dependence of the $y$ component of the electric polarization $ P_y $ (b).}
	\label{fig_t1}
\end{figure}

The  total free-energy  density receives contributions from the exchange interaction $ A_{ex} (\nabla \vec{m})^2 $, the Zeeman energy $ -\vec{H}_{\text{ext}} \cdot \vec{M} $, magnetic anisotropies, and magnetoelectric contribution is the presence of an applied electric field that couples to the ferroelectric polarization. The latter results in the energy contribution $\int \vec{E\big(\vec{r}\big)}\vec{P\big(\vec{r}\big)}d^{3}\vec{r}$.  A switching off the electric field the system relaxes to the pure magnetic configuration. Here we did not discuss the dissipation of the ferroelectric energy nor the dynamics of the  polarization (as a separate order parameter) and its damping, which was presented in a recent work \cite{seyyed_damping}. We  assume that such dissipation related to the fluctuation in the polarization is accounted for by parts of the Gilbert damping. We recall that  ferroelectric energy dissipation goes along with a diminishing spin current.  Thus, generating the thermal magnonic spin currents (as shown below) comes at the cost of spending electric work. For illustrating the relaxation process we present in Fig. \ref{fig_t1} the effective polarization  after switching-off the electric field. The ratio between the ferroelectric energy and the relaxation time $\Delta Q=\frac{1}{\tau}\int \vec{E\big(\vec{r}\big)}\vec{P\big(\vec{r}\big)}d^{3}\vec{r}$ can be viewed as a qualitative measure for the energy dissipation rate. 

\section{Results and interpretations}
\label{part4}

Applying  at $T = 15$K an $E$ field only to a part of the chain ($ 0 \leq x \leq 900 $ nm)  (Fig. \ref{fig_1}(a)) results in a finite $ J_{ex} $ with a steady-state character reversing sign when $E_{y}$ is reversed.  The magnitude  $ |J_{ex}| $ is not equal for $\pm E_{y}$ due to the spin wave dispersion relations (for technicalities and materials parameters, please see Appendix A). No spin current is generated if $E_{y}=0$. Fig. \ref{fig_1}(c) exposes the asymmetry of $ J_{ex} $ vs.~the magnetization angle $ \phi $. Reversing both $\vec{m}$ and $\vec{E}$ leaves $ J_{ex} $ unchanged.
The spin current also diffuses into the $E_{y}-$ field-free part covering the whole chain. One can also evaluate the stochastically averaged effective DM strength  associated with the spin currents in Fig. \ref{fig_1}(a).

\begin{figure}
	\centering
	\includegraphics[width=0.45\textwidth]{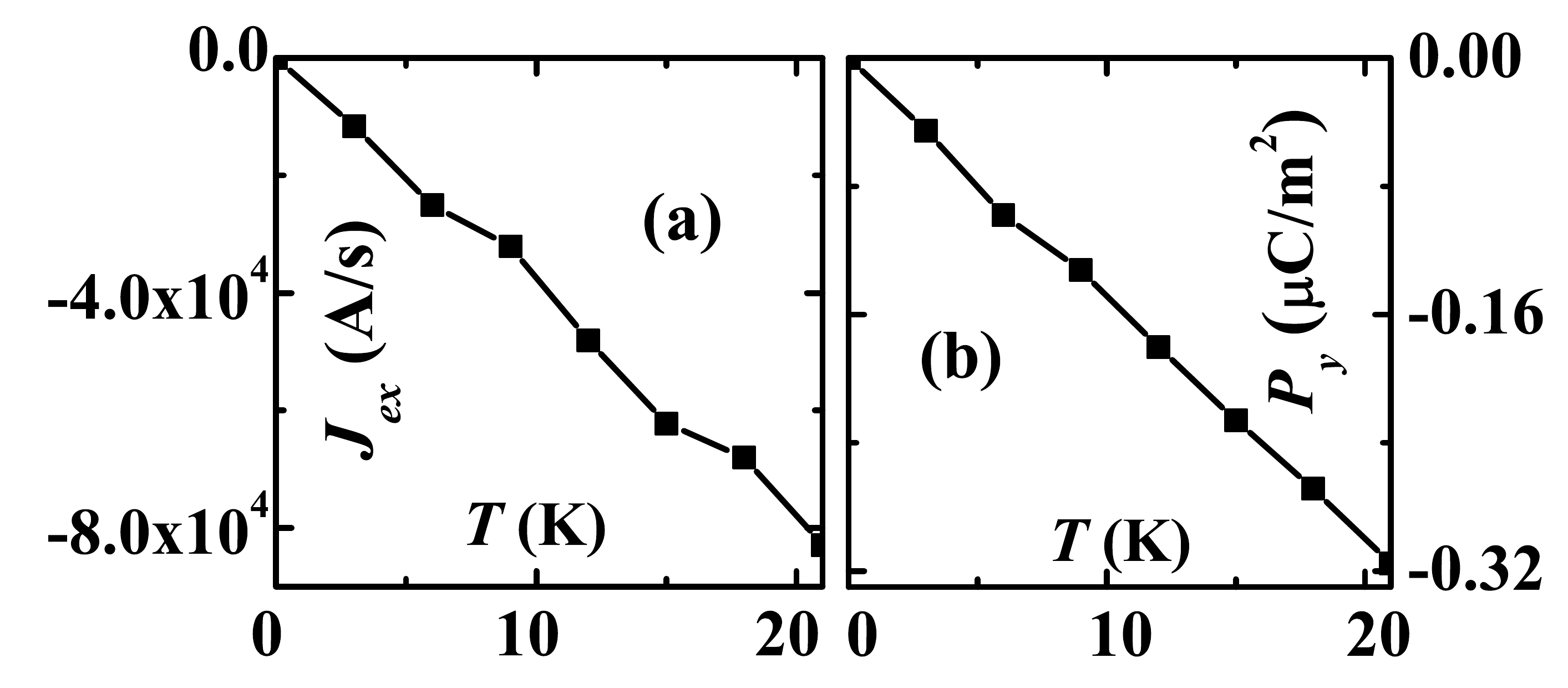}
	\caption{(a) Averaged exchange spin current $ J_{ex} $ in the region ($ 1000 \text{ nm} \leq x \leq 2000 $ nm ) as a function of a uniform  $T$. (b) The ensemble averaged effective polarization of the whole system as a function of \textit{T}. Here,  $\phi = \pi / 2 $ and $ E_{y}=-0.05 $ MV/cm in the left subsystem $ 0 \leq x \leq 900 $ nm. }
	\label{fig_3}
\end{figure}

We can introduce a local "effective  magnonic temperature" as inferred  from the local magnon density.  This additional entropic contribution to the free energy density enforces the magnons to diffuse on a length scale given by the \textit{effective} magnonic temperature profile. On the length scale of the considered chain the formed effective magnonic temperature profile is close to linear and so is the behavior of the diffusing spin current. This  (quasi linear) behavior persists when increasing the applied uniform temperature  (cf.  Fig. \ref{fig_3}).   Being  thermal fluctuation driven, $ J_{ex} $  increases when increasing $T$. Summarizing this effect, an $E$-field gradient may convert thermal fluctuations to directed magnonic spin current whereby spin relaxation  quantified by the Gilbert damping is also essential. What is the physics and the systematics behind the  values and the directions of $ J_{ex} $ observed in Figs.~\ref{fig_1}-\ref{fig_3}?

For a further  insight we  inspect the spectral characteristics of the participating magnons while restricting the discussion to the small-amplitude spin waves that propagate along the \textit{x} axis:
$ \vec{m}_{sw} = (m_{\theta} \vec{e}_{\theta} + m_{\phi}\vec{e}_{\phi}) e^{i(\pm k_{\pm}x - \omega t)} $, where $ \omega $ is the magnon frequency, $ k_{\pm} $ are the magnon wave-vectors, and $ (\pm) $ indicates that the magnon propagation direction is parallel or antiparallel to the  $ \pm x $ axis.
With the ansatz $ \vec{M}=\vec{M}_{0} + \vec{m}_{sw}$ for the solution of the
LLG equation we deduce for the  magnon dispersion the relation \cite{Tliu20}
\begin{equation}
\displaystyle \omega = \gamma H_{0} + \frac{2 \gamma A_{ex}}{\mu_{0} M_{s}}k_{\pm}^{2} \pm\frac{2 \gamma c_{E} E_{y}}{\mu_{0} M_{s}} \text{sin}\theta\text{sin}\phi k_{\pm}.
\label{eq_2}
\end{equation}
For $ \theta = \pi/2 $ and the \textbf{}$\omega$ defined from Eq.(\ref{eq_2}), the difference between the wave vectors is $ k_{+} -k_{-} = -c_{E} E_{y} \mathrm{sin}\phi / A_{ex} $
The explicit expressions for $ k_{+},~k_{-}$ deduced from Eq.(\ref{eq_2}) and read $ k_\pm^{\mathrm{left}} = k_\pm = (2 A_{ex})^{-1} \Big[\mp c_E E_y \sin \theta \sin \phi + \Big((c_E E_y \sin \theta \sin \phi)^2 - 4 \mu_0 M_s A_{ex} H_0 + 4 \mu_0 M_s A_{ex} \omega / \gamma \Big)^{1/2}\Big]$, and $ k_\pm^{\mathrm{right}} = k_0 = A_{ex}^{-1}\sqrt{- \mu_0 M_s A_{ex} H_0 + \mu_0 M_s A_{ex} \omega / \gamma} $.

The following scenario for the mechanism  of a spin current generation emerges:
At a finite uniform $T$ the magnons diffuse equally in  both directions $ \pm x $.An applied inhomogeneous static $E_{y}$ couples to the system as an emergent DM interaction. This effect is not quadratic in $E_{y}$
(as for Raman-type processes) as it is  primarily induced by thermal fluctuations, not by  $E_{y}$.
Once the DM interaction is stabilized in a steady state, the velocities of  magnons propagating in  opposite $x$ directions become different.
The phenomenological expression for the magnonic spin current reads
\begin{equation}
\displaystyle J_{ex} = \int [n_{+}(\omega)v_{ex}^{+}(\omega)- n_{-}(\omega)v_{ex}^{-}(\omega)] (-\hbar) d\omega.
\label{eq_8}
\end{equation}
The densities of the magnons $n_{\pm}$ propagating in the different directions  are proportional to the amplitudes $\rho^{2}_{\pm}$ of the corresponding spin waves $\int \rho_{\pm}(\omega)e^{\pm i k_{\pm}(\omega) x} d\omega$. The group velocity of the exchange magnons follows from the dispersion relation Eq.(\ref{eq_2})  as  $v_{ex}^{\pm}=\frac{\partial \big(\frac{2 \gamma A_{ex}}{\mu_{0} M_{s}}k_{\pm}^{2}\big)}{\partial k_{\pm}}=\frac{4 \gamma A_{ex}}{\mu_{0} M_{s}} k_{\pm} $. For a uniform $T$ and $E_{y}=0$ the group velocities of the right and left propagating magnons are equal $ k_{+} = k_{-}=k_{0}$, and the same applies to the magnon densities $ n_{+} = n_{-}=n_{0}$.
As a result, the spin current is zero $ J_{ex}=0$. Applying an electric field to a part of the system (say the left one) we infer for the steady state $ k_{+} -k_{-} = -c_{E} E_{y} \mathrm{sin}\phi / A_{ex} $. According to Eq. (\ref{eq_8}) a nonzero magnonic current $ J_{ex} $ emerges in the left part. When $ E_{y} > 0 $ and $ \phi = \pi / 2 $, $ k_{+} -k_{-} < 0 $ the spin current $ J_{ex} $ is positive in the left part, as shown in Fig. \ref{fig_1}. For the induced magnonic spin current in the right part (where $E_{y}=0$),  interfacial effects  are important (interfacial regions are those where the $E$ field gradient is finite).

\section{Interfacial effects}
\label{part5}

At a uniform temperature and in absence of electrical field gradients the  thermally activated  magnons in our system propagate equally  in both $ \pm x $ directions resulting in a zero net spin current.
Now let us consider   the magnons propagating in the $ + x $ direction, i.e.  form the left part where $ E_y \ne 0 $ to the right part where $ E_y = 0 $.
The region where $ E_y $ is inhomogeneous, i.e. where it changes from $ E_y \ne 0 $ to $ E_y = 0 $ we call for simplicity  "interfacial region", even though there is no physically dividing interface in the sample.   Obviously, the region in which the electric field changes is determined by screening charges. Hence this region is very narrow on the typical scale of the magnon wave length. The  magnonic  dynamic is affected
through  changes in the DM interaction, i.e.  in the region of spatially varying $E$ field.
To uncover the magnon dynamics in the presence of the "interfacial region"  we note that when  passing through the interface region, the magnon energy   is conserved.
Due to the difference in the dispersion relations (Eq. (2)), the magnon frequencies are equal $\omega \big(k_{\pm},E\neq 0\big)=\omega \big(k_{0},E= 0\big)$ only if $k_{\pm}\neq k_{0}$.
On the other hand the equations of motion dictate a continuous and smooth transitions across the interfacial region (and in the whole structure)  of   left-to-right ($ F_{sw}$) or right-to-left  propagating spin waves ($ G_{sw}$). Without loss of generality we assume the interfacial region to be centered at $x=0$.
As this region is extremely sharp on  the scale of the wave-length of spin waves, continuity and smoothness dictate that
\begin{eqnarray}
\displaystyle F^{in}_{sw}\big(x=0\big)=F^{out}_{sw}\big(x=0\big),
\nonumber \\
\frac{F^{in}_{sw}\big(x=0\big)}{\partial x}=\frac{F^{out}_{sw}\big(x=0\big)}{\partial x},
\label{smoothness1}
\end{eqnarray}

\begin{eqnarray}
\displaystyle G^{in}_{sw}\big(x=0\big)=G^{out}_{sw}\big(x=0\big),
\nonumber \\
\frac{G^{in}_{sw}\big(x=0\big)}{\partial x}=\frac{G^{out}_{sw}\big(x=0\big)}{\partial x}.
\label{smoothness2}
\end{eqnarray}
Here $F^{in}_{sw}$ and $ F^{out}_{sw}$ are from left-to-right propagating
spin waves in the regions left and right to the interface, respectively. Similarly, $G^{in}_{sw}$ and $G^{out}_{sw}$
are from right-to-left propagating waves in the right and left regions, respectively. Eqs. (\ref{smoothness1}) and (\ref{smoothness2})  are just Fresnel equations for scattering of continuous waves \cite{Hecht2002,Stancil2009}. The interfacial region acts as a sharp scatterer for the spin waves. Considering  the smooth continuity criteria Eqs.~(\ref{smoothness1},\ref{smoothness2}) we deduce the transmission and reflection coefficients:
Due to the magnon reflection at the interface, i.e. the region where the electric field is inhomogeneous, the spin waves at the left side of the interface are described as
\begin{equation}
F^{in}_{sw} = \int (\rho_0 e^{ik_+ x - i\omega t} + R_1 \rho_0 e^{-ik_- x - i\omega t}) d\omega,
\end{equation}
where $ k_\pm $ are the magnon vectors in the $ \pm $ direction in the left side, $ \rho_0 $ is the amplitude of the incoming spin waves, and $ R_1 $ is the ratio of the reflected spin waves. The magnons at the right side are
\begin{equation}  F^{out}_{sw} = \int (S_1 \rho_0 e^{ik_0 x - i\omega t}) d\omega ,\end{equation}
$ S_1 $ is the ratio of the transmitted spin waves. The wave vectors $ k_\pm $ and $ k_0 $ are determined by Eq. (2) when $ E_y \ne 0 $ and $ E_y = 0 $, respectively. Thus, the magnon dispersions  in different parts are different, yet the continuity condition for spin waves through the interfacial region  holds true. From the continuity and the smoothness of the magnon propagation   we infer the ratio between the transmitted and the reflected magnons to be
$$ S_1 = \frac{k_+ + k_- }{k_0 + k_-};\quad R_1 = \frac{k_+ - k_0 }{k_- + k_0} .$$   Similarly, we find
for  magnons propagating from the right part to the left part
\begin{equation} G^{in}_{sw} = \int (\rho_0 e^{-ik_0 x - i\omega t} + R_2 \rho_0 e^{ik_0 x - i\omega t}) d\omega,\end{equation}
and
\begin{equation} G^{out}_{sw} = \int (S_2 \rho_0 e^{-ik_- x - i\omega t}) d\omega .\end{equation}
The transmission ratio is $$ S_2 = \frac{2 k_0}{ k_0 + k_-}  ,$$ and the reflection ratio is $$ R_2 = \frac{k_0 - k_-}{k_0 + k_-} .$$  We note that the interactions between the magnons with different frequencies are neglected in the above considerations.

The exchange spin current at the right side of the interface $ J_{ex} $   receives contributions from  a) the magnons transmitted from the left part through the interface towards the right part in the $ +x $ direction,   b) the incoming magnons from the right part that propagate towards the left part in the $ -x $ direction, and c) magnons reflected  in the right part. Hence we have $ J_{ex}=\int \frac{4 \gamma A_{ex}}{\mu_{0} M_{s}} [(S_{1} + R_{2} )^2 - 1] n_{0} v_{0}(-\hbar) d\omega $. Here $ v_{0} = \frac{4 \gamma A_{ex}}{\mu_{0} M_{s}} k_{0} $ is the group velocity of the exchange magnons in the right part and $ n_{0} $ is the propagating magnon density.

Considering a finite $T$ and $E_{y} \ne 0$ in the left part,
for $ \theta = \pi /2 $, we find $ S_{1}+R_{2}+1 $ is always positive, and $ S_{1}+R_{2}-1 = -\frac{c_{E} E_{y} \mathrm{sin} \phi / A_{ex}}{k_{0}+k_{-}}$.   $ k_{-} $ is the wave vector of magnons in the left part with $E_{y} \ne 0$ that propagate along the $ -x $ direction. When $ E_{y} >0 $ and $ \phi = \pi / 2 $, the coefficient  $ S_{1}+R_{2}-1 $ is negative and  $ J_{ex} $ is positive. This result is confirmed by numerical calculations see Fig. \ref{fig_1}. For $ E_{y} <0 $, the coefficient $ S_{1}+R_{2}-1 $ turns positive and  the sign of the spin current $ J_{ex} $ is reversed. Besides, since $ k_{-} $ is smaller for negative effective DM the constant $ E_{y} <0 $, and  the corresponding spin current density is larger compared to the case $ E_{y} >0 $. This asymmetry feature is also confirmed numerically (Fig. \ref{fig_1}).  The dependence of the spin current $ J_{ex} $ on the angle $ \phi $ is defined by the term $-\frac{c_{E} E_{y} \mathrm{sin} \phi / A_{ex}}{k_{0}+k_{-}}$ (cf.~Fig. \ref{fig_1}). The asymmetry with respect to $ \phi $ appears  because $ k_{-} $ is smaller for positive $ \phi $.  An increase in the uniform temperature enhances the  magnon density \textit{n} and the spin current $ J_{ex}^{z} $ (see Fig. \ref{fig_3}).
If the electric field is applied to the left part the transmission ratio of magnons in the left and right parts are $ S_{1} =\frac{k_{+}+k_{-}}{k_{-}+k_{0}} $, and $ S_{2} =\frac{2 k_{0}}{k_{-}+k_{0}} $. For a negative DM constant and $ \phi = \pi /2 $ we have $ E_{y} <0 $, $ k_{-} < k_{0} < k_{+} $. Surprisingly both transmission ratios are larger than one $ S_{1,2}>1$, meaning  that the electric field enhances the magnon density at the interface. A change of the sign of the electric field ($ E_{y} > 0$) decreases the magnon density. Using the ansatz of the transmitted and reflected waves we estimate
the magnon density in the vicinity of interface from the right and from the left to be
\begin{eqnarray}
\displaystyle n_{r}=\frac{\big[\big(S_{1}+R_{2}\big)^{2}+1\big]\rho_{0}^{2}}{2g\mu_{B}M_{s}}.
\label{density1}
\end{eqnarray}
and
\begin{eqnarray}
\displaystyle n_{l}=\frac{\big[\big(S_{2}+R_{1}\big)^{2}+1\big]\rho_{0}^{2}}{2g\mu_{B}M_{s}}.
\label{density2}
\end{eqnarray}
Here $\frac{2\rho_{0}^{2}}{2g\mu_{B}M_{s}}$ is the initial magnon density. As we see from Eq. (\ref{density1}) the magnon density  increases when
$\big[\big(S_{1}+R_{2}\big)^{2}+1\big]>2,~~\big[\big(S_{2}+R_{1}\big)^{2}+1\big]>2$, and decreases when
$\big[\big(S_{1}+R_{2}\big)^{2}+1\big]<2,~~\big[\big(S_{2}+R_{1}\big)^{2}+1\big]<2$.

\begin{figure}
	\centering
	\includegraphics[width=0.46\textwidth]{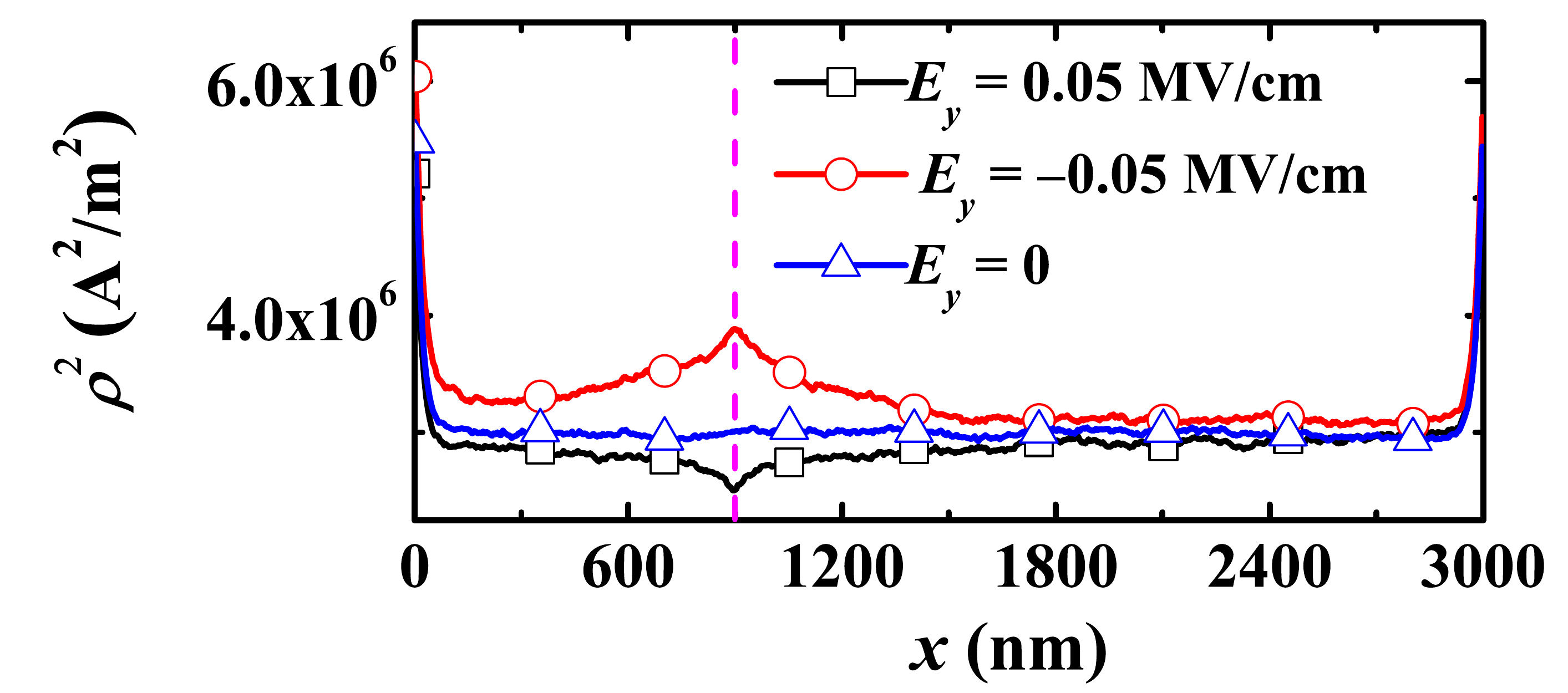}
	\caption{The magnon local density $ \rho^2 $ for the electric fields $ E_{y}=0 $   (blue triangles), $ E_{y}=0.05 $ MV/cm (black squares), and $ E_{y}=-0.05 $ MV/cm (red circles).}
	\label{fig_4}
\end{figure}

To check the viability of the analytical estimations, and to quantify the local density of magnons, in micromagnetic simulations we utilize the standard definition of the magnon density $ \rho^2 = \langle M_{\theta}^{2} + M_{\phi}^{2} \rangle $. The spatial distribution of magnon density $ \rho^2 $ is shown in Fig. \ref{fig_4}, evidencing both effects: an enhancement ($ E_{y} <0 $, $ \phi = \pi /2 $) and a depletion ($ E_{y} >0 $, $ \phi = \pi /2 $ ) of the local magnon densities.
$ k_{+} $ is smaller for $ E_{y} > 0 $, implying a stronger enhancement  associated with  magnons flowing  from the interface (negative $ J_{ex} $). The reduction effect corresponds to a positive $ J_{ex} $. This trend is preserved for other values of $ \phi $.

\begin{figure}
	\centering
	\includegraphics[width=0.49\textwidth]{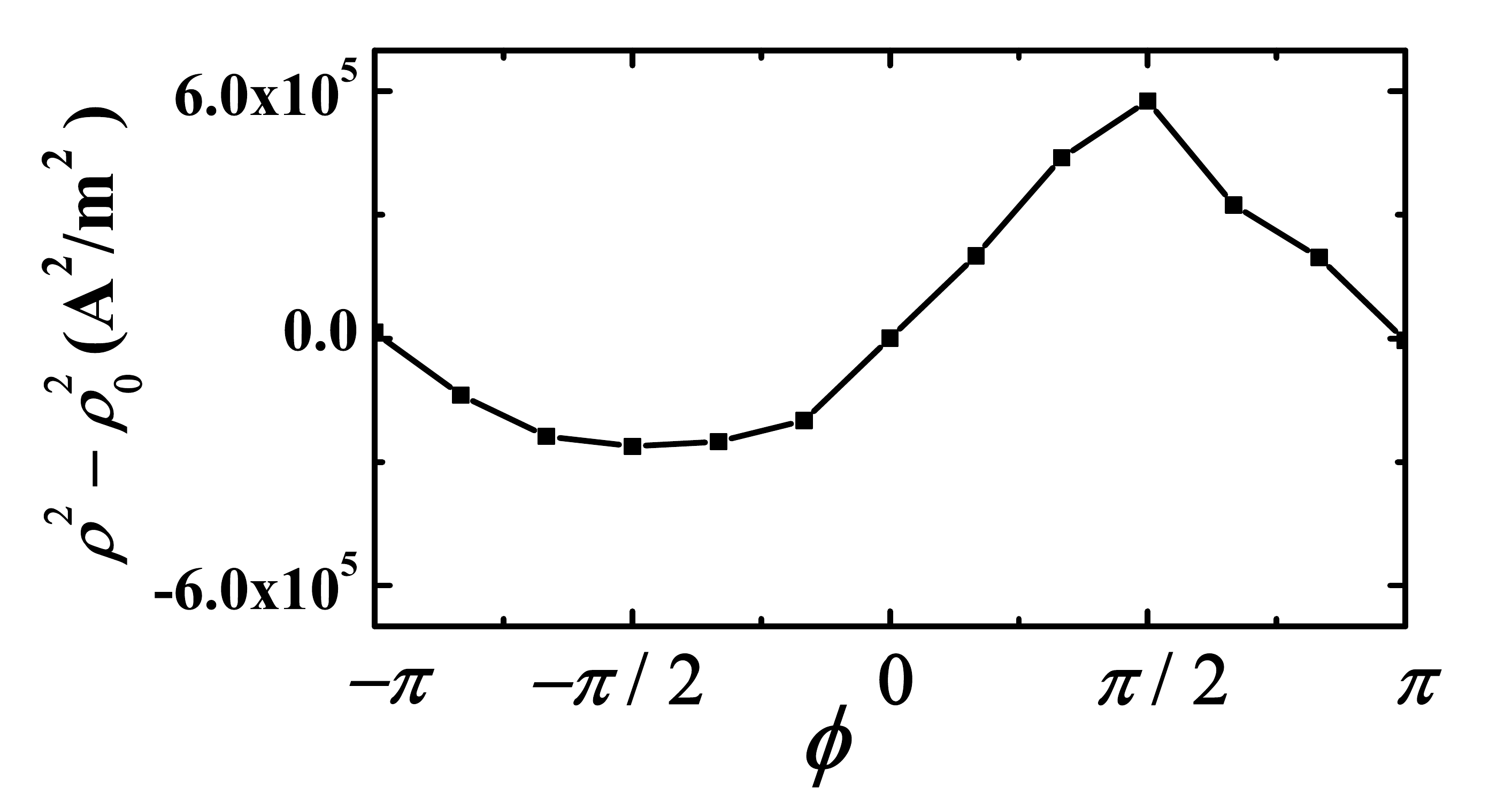}
	\caption{Averaged local magnon density  variation $ \rho^2 - \rho_0^2 $ in the region of $ 600 \text{ nm} \leq x \leq 1200 \text{ nm} $ as a function of  $\phi $ for $ E_y = -0.05 $ MV/cm. }
	\label{fig_S1}
\end{figure}

Fig. \ref{fig_S1} shows the variation of the averaged local magnon density $ \rho^2 - \rho_0^2 $ as a function of $ \phi $, where $ \rho_0^2 $ is the magnon density in the absence of the electric field ($ E_y = 0 $). The dependence of the density variation on the angle $ \phi $ manifests a non-monotonic and an asymmetric behavior. We see an enhancement ($ \rho^2 - \rho_0^2 > 0  $) for $ 0 < \phi < \pi  $ and a reduction ($ \rho^2 - \rho_0^2 < 0  $) for $ -\pi < \phi < 0 $. The magnon density variation has a maximum for $ \phi = \pm \pi / 2 $, and for $ \phi = \pi / 2 $ the maximum is slightly larger.

Notably, in spin caloritronics spin currents in ferromagnetic insulators are occasionally  discussed in connection with a nonuniform magnon chemical potential \cite{JBar12,JBar1201,Xiao}. Here we argue by means of the nonuniform magnon density that we can access and analyze directly with our numerical and analytical methods. The connection to the chemical potential formulation can be found in Ref. \cite{Xiao}.

\begin{figure}
	\centering
	\includegraphics[width=0.49\textwidth]{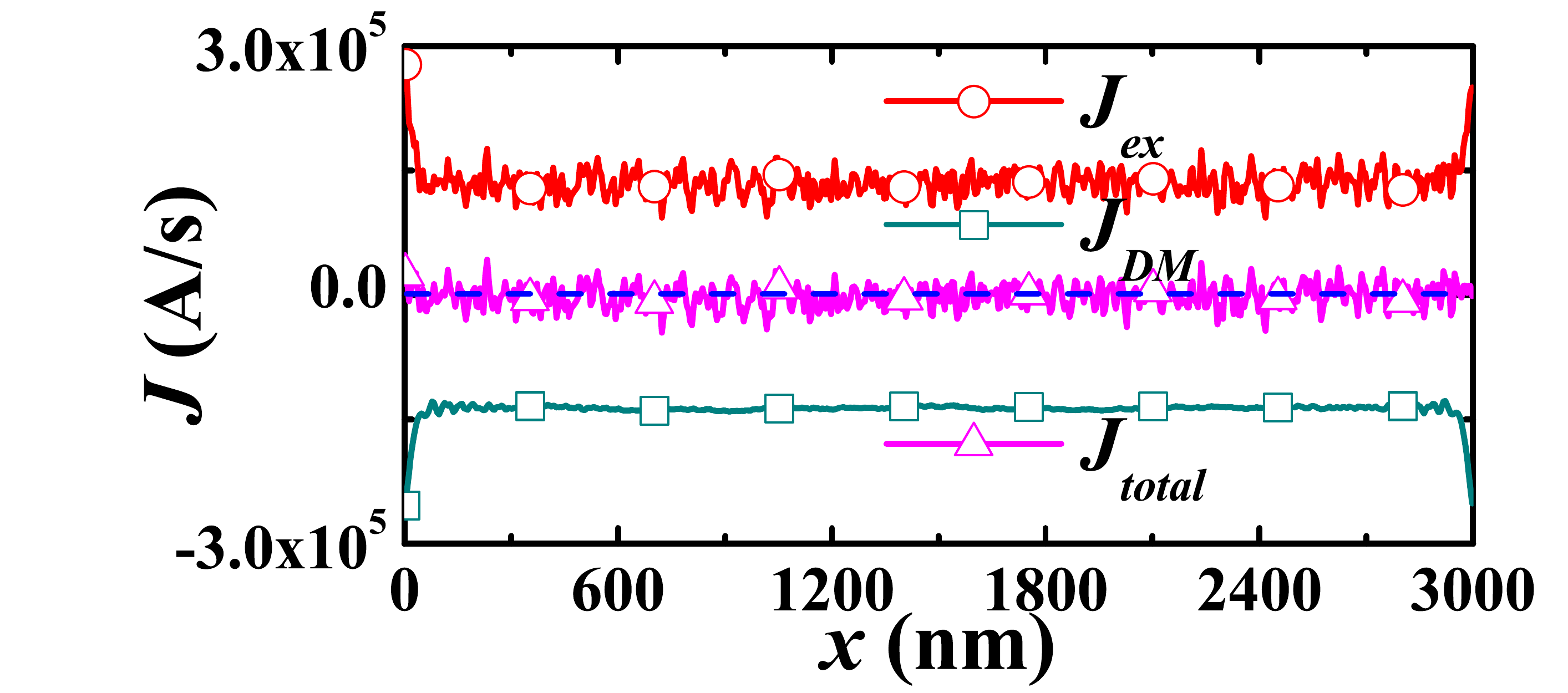}	
	\caption{ Profiles of the exchange $ J_{ex} $ (red circles), the chiral $ J_{D} $ (black squares) and the total $ J_{\text{total}}$ (blue triangles) the magnonic spin current. Electric field $ E_{y} $ is applied to the whole system with $ E_{y}= 0.05 $ MV/cm and $ T = 15 $ K. }
	\label{fig_S6}
\end{figure}

\begin{figure}
	\centering
	\includegraphics[width=0.45\textwidth]{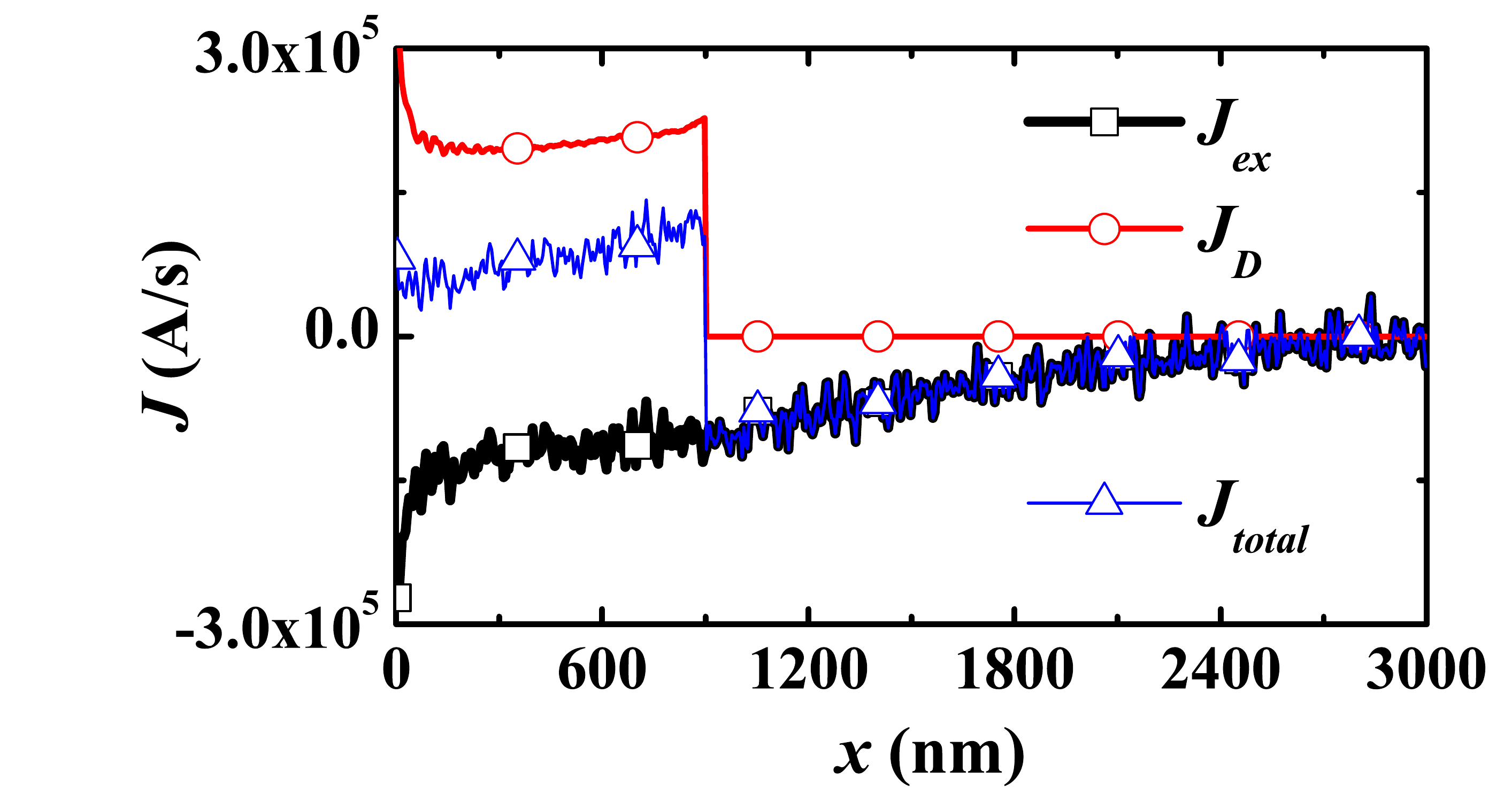}
	\caption{The exchange $ J_{ex} $ (black squares), the chiral $ J_{D} $ (red circles), and the total $ J_{\text{total}}$ (blue triangles) thermally averaged magnonic spin currents.
		$ E_{y} $ is applied to the left subsystem $ 0 \leq x \leq 900 $ nm with $ E_{y}=-0.05 $ MV/cm. }
	\label{fig_5}
\end{figure}

Using the expressions for the chiral spin current $J_{D}$ and the exchange spin currents $J_{ex}$ (see Eq.(\ref{eq_5}), Eq.(\ref{chiralcurrent})), we calculate the total spin current $ J_{\text{total}} = J_{ex} + J_{D} $.
When a uniform electric field is applied to the whole system, the DM spin current compensates for the exchange spin current and the total spin current vanishes Fig. \ref{fig_S6}.
When the electric field is applied to the left subsystem only, the DM spin current vanishes in the right subsystem, as shown in Fig. \ref{fig_5}. The exchange spin current is equal to the total spin current in the right part, and it is smaller than the DM spin current in the left part.
The total magnonic spin current induced in the left part  is opposite to the current in the right part of the system. A positive spin current in the left part evidences a magnon  flow away from the interface.
\begin{figure}
	\centering
	\includegraphics[width=0.45\textwidth]{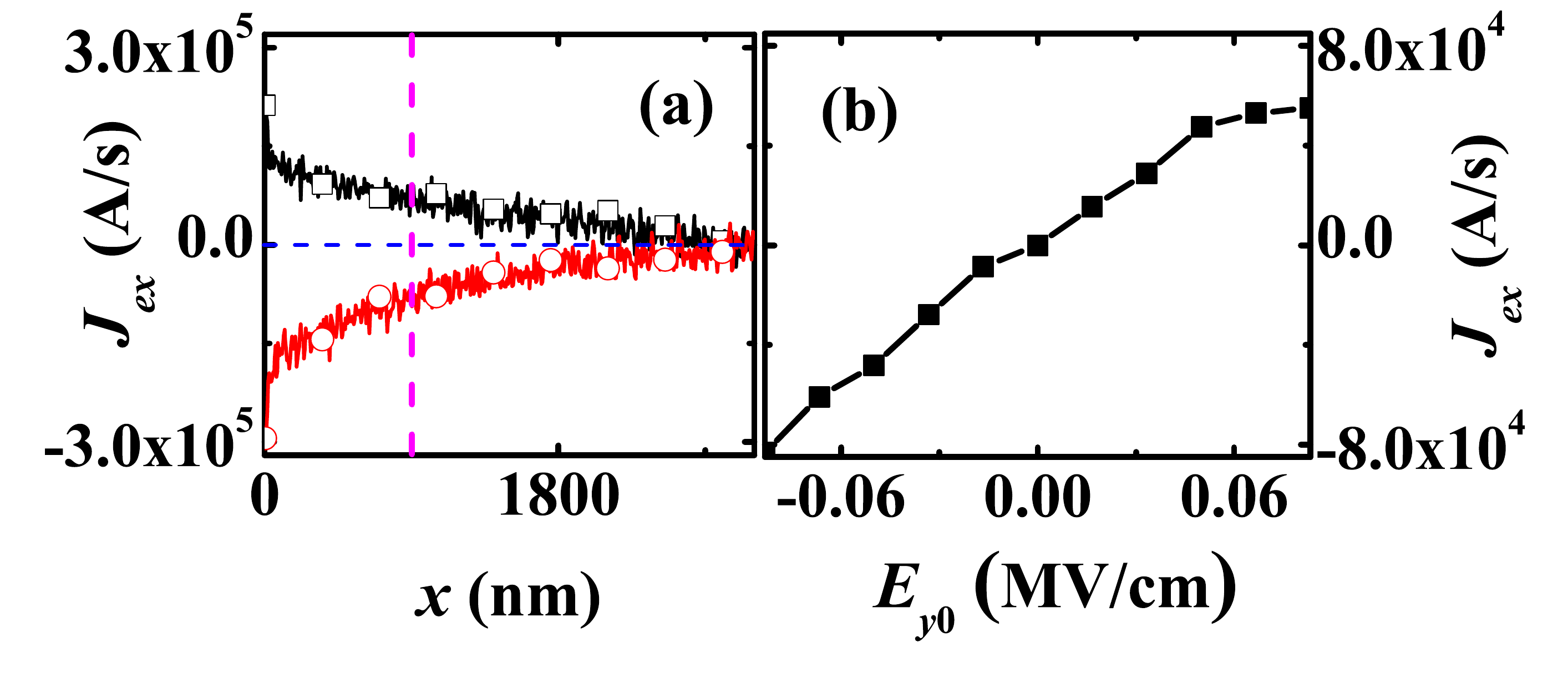}
	\caption{(a) The exchange spin current $ J_{ex} $, when an electric field gradient  $ E_{y}=E_{y0}\frac{900 \text{nm} - x}{900 \text{nm}} $,  $ E_{y0}=  0.05 $ MV/cm (black squares) and $ E_{y0}=  -0.05 $ MV/cm (red circles) is applied to the left subsystem. (b) The thermally averaged $ J_{ex} $ in the region of ($ 1000 \text{ nm} \leq x \leq 2000 $ nm ) as a function of $ E_{y0} $.}
	\label{fig_6}
\end{figure}
A finite steady-state spin current requires an inhomogeneous electric field.
This is realized by applying a homogeneous electric fields to part of the system only. Homogeneous electric fields acting across  the whole system results in a  zero  spin current.

The DM spin current is quantified as
\begin{equation}
\displaystyle J_{D}=\int [n_{+}(\omega)v_{\text{DM}}^{+}(\omega)- n_{-}(\omega)v_{\text{DM}}^{-}(\omega)] (-\hbar) d\omega.
\label{eq_9}
\end{equation}
The magnons velocity  related to the chiral interaction follows from the dispersion  $ v_{\text{DM}}^{\pm} = \pm\frac{2 \gamma c_{E} E_{y}\sin\phi}{\mu_{0} M_{s}} $. For a uniform temperature profile the densities of magnons diffusing to the left and to the  right are equal $ n_{+} = n_{-} $. Therefore, if the electric field is applied along the whole system from Eqs. (\ref{eq_8}) and (\ref{eq_9}) follows that  the total spin current is zero, $ J_{D} + J_{ex} = 0  $. The zero current for the homogeneous electric field applied on the whole system is also testified by our full numerical simulations.

To explore the effect of nonuniform electric field, let us consider the gradient of the electric field  $ E_{y}=E_{y0}\frac{900 \text{nm} - x}{900 \text{nm}}  $ applied on the left part of the system $ 0 \text{ nm} \leq x \leq 900 \text{ nm} $, while in the right subsystem $ E_{y}=0 $. For $ \phi = \pi /2 $, the exchange magnonic spin current is depicted  in Fig. \ref{fig_6}(a) confirming that the magnonic spin current $ J_{ex} $ in the right subsystem is negative for $ E_{y}<0 $, and is also negative in the left subsystem. The gradient of the electric field leads to a gradient in the spin current $ J_{ex} $. The larger the applied field the larger is the induced current,  as demonstrated by  Fig. \ref{fig_6}(b).

A key issue for the generation of the spin current is the relative  orientation of the external electric field and the equilibrium ground state magnetization.
For instance, $E_y$ with $\phi = 0,\pi$ results in zero current while for a different $\phi$ the current can be finite.
We note the link between the effective polarization and the exchange magnonic spin current $\vec{P} = \frac{c_{E} \mathrm{sin} \theta J_{ex}}{l_{ex} M_s^2} \vec{m}_0 \times \vec{e}_{ex}$. $ l_{ex} = \frac{2 \gamma A_{ex}}{\mu_{0} M_{s}^{2}} $ and the vectors $ \vec{m}_0 $,  $\vec{e}_{ex}$ set the direction along which the magnonic spin current flows
(\textit{x} axis in our case) . Both $ \vec{P} $ and $ J_{ex} $ are emergent phenomena, they vanish  in the ground state and appear for a non-uniform $\vec E$ field and finite uniform $T$.
The polarization (magnonic spin current) is finite  when $\vec E$ has a component parallel to
$ \vec{m}_0 \times \vec{e}_{ex} $. For finite $ E_z $   and $ \vec{m}_0 $ being directed towards \textit{y} direction ($ \phi = 0, \pm\pi $) we predict a finite spin current.

\section{Effect of different electric field components}
\label{part6}

\begin{figure}
	\centering
	\includegraphics[width=0.49\textwidth]{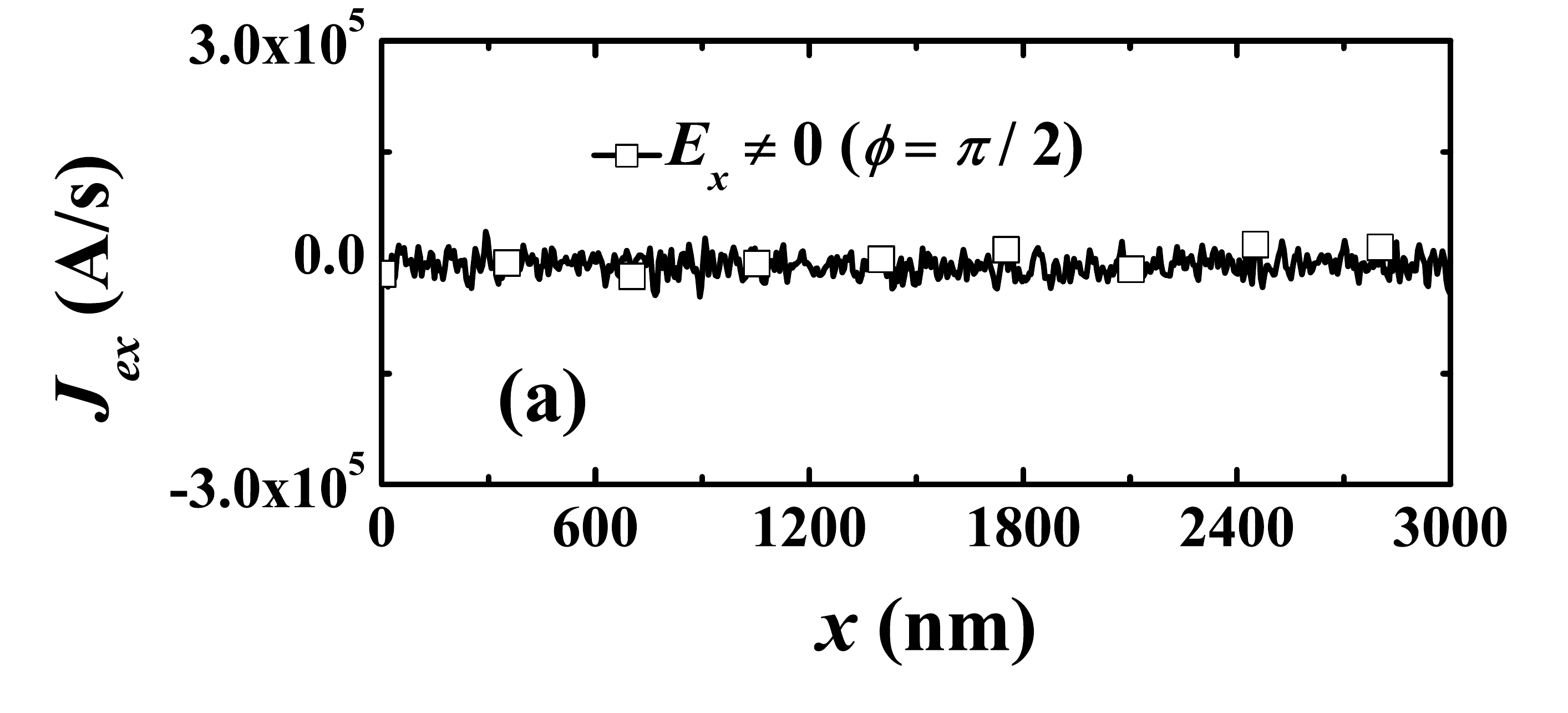}
	\includegraphics[width=0.49\textwidth]{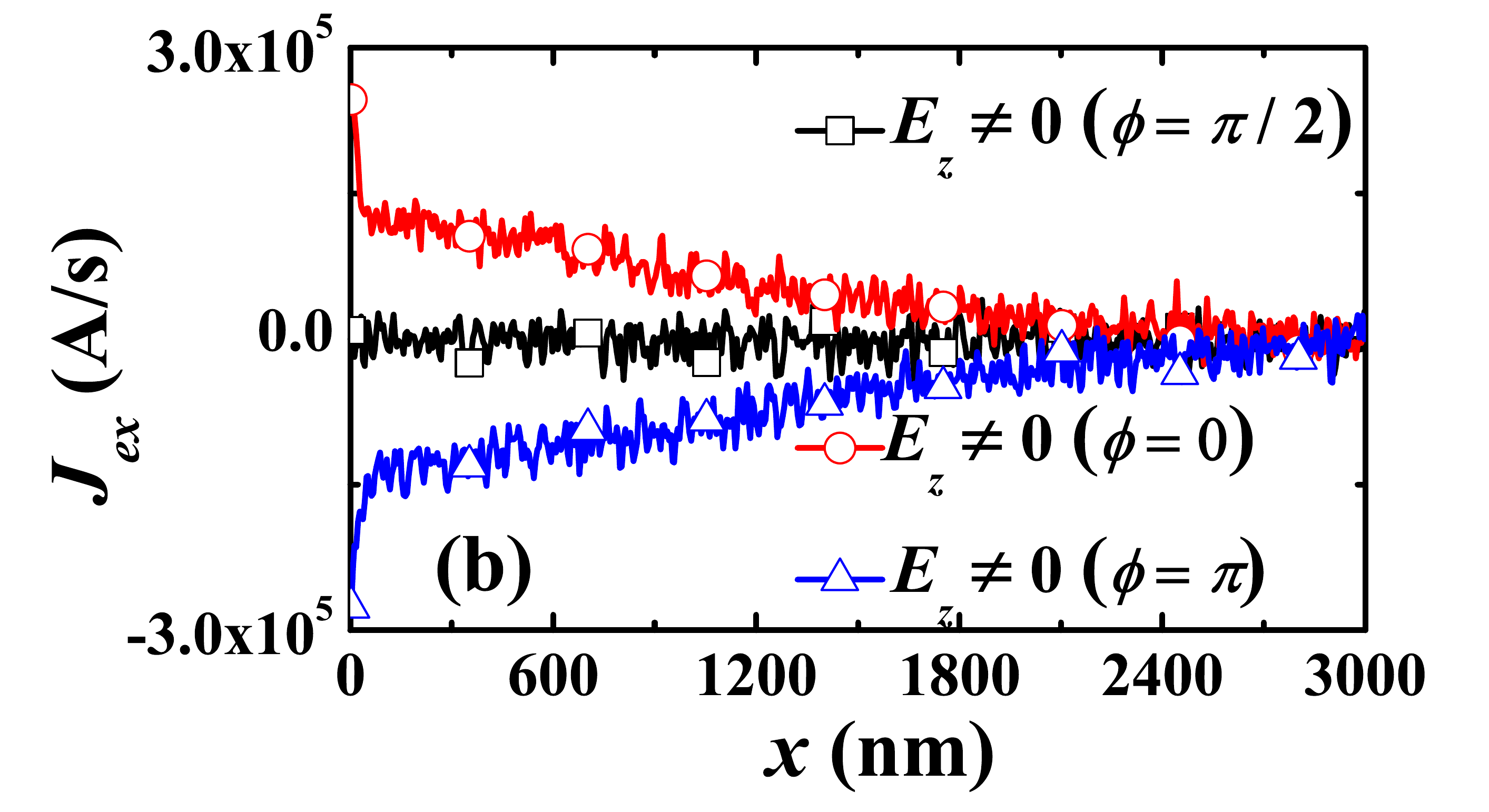}
	\caption{(a) Profile of the exchange spin current $ J_{ex} $. The electric field $ E_{x}=-0.05 $ MV/cm (a) and $ E_{z}=-0.05 $ MV/cm (b) is applied to the left subsystem $ 0 \leq x \leq 900 $ nm. The magnetization angle are $ \theta = \pi / 2 $ and $ \phi = 0 $ (blue triangles), $ \pi / 2 $ (black squares) and $ \pi $ (red circles).}
	\label{fig_S2}
\end{figure}

\begin{figure}
	\centering
	\includegraphics[width=0.49\textwidth]{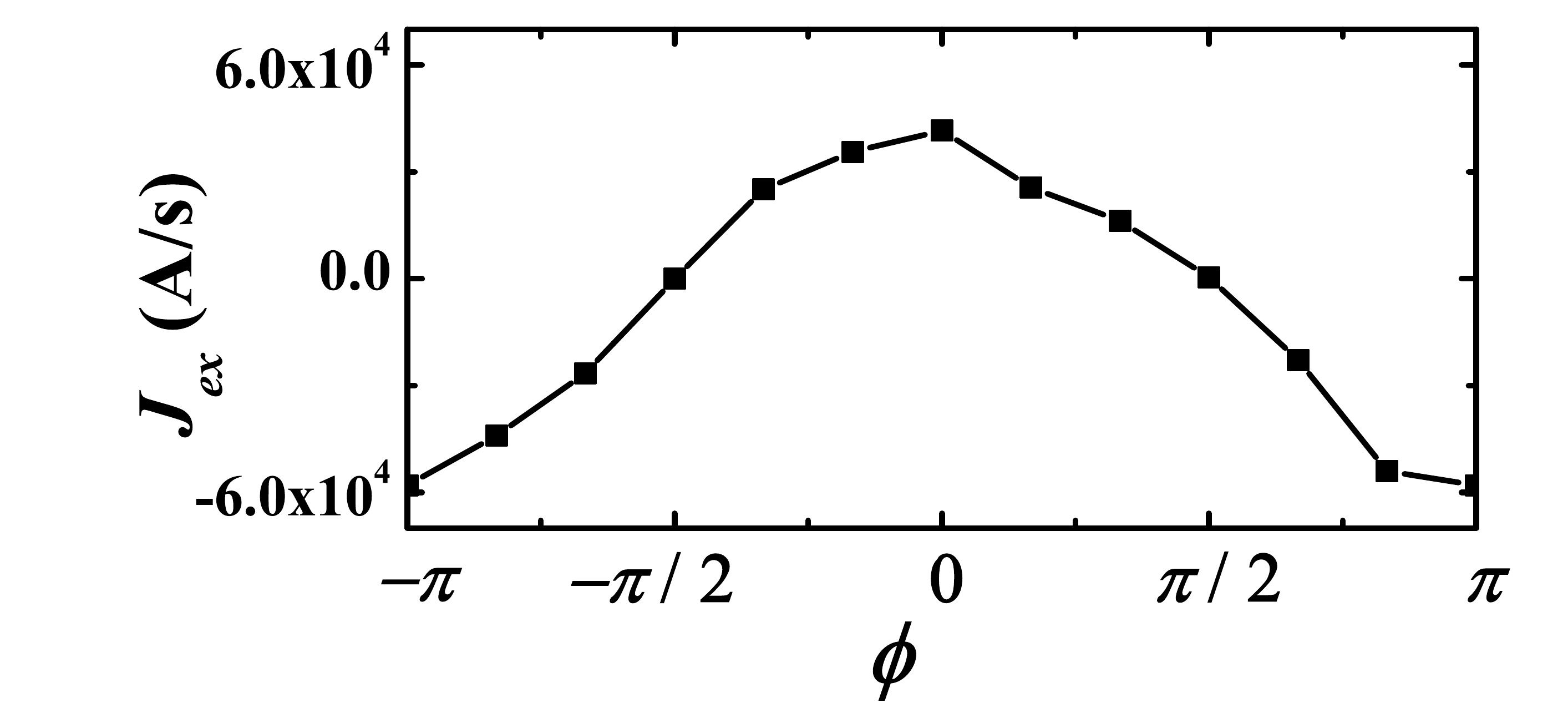}
	\caption{The averaged exchange spin current $ J_{ex} $ in the region of $ 1000 \text{ nm} \leq x \leq 2000 \text{nm} $ as a function of  $\phi $ for $ E_z = -0.05 $ MV/cm. }
	\label{fig_S3}
\end{figure}

Fig. \ref{fig_S2} shows the exchange magnonic spin current for  different configurations of the applied electric field. An electric field $ E_x$ applied along the $ x $ axis  does not contribute to the spin current $ J_{ex}=0$. In the case  an electric field $ E_z$ along the $ z $ axis the  spin current is still zero $ J_{ex}=0$ for $ \theta = \pi / 2 $, $ \phi = \pi / 2 $, while the current is positive (negative) for $ \theta = \pi / 2 $ and $ \phi = 0,\pi $. The connection between $ J_{ex} $ and the magnetization angle $ \phi $ is shown in Fig. \ref{fig_S3}, for $ \theta = \pi / 2 $ and $ E_z = -0.05 $ MV/cm. The dependence of the magnonic current $ J_{ex} $ on the angle $ \phi $ manifests a non-monotonic and an asymmetric behavior.  The spin current induced by the $ E_z $ component has a maximum for $ \phi = 0 $ and $ \phi = \pm \pi $, and for $ \phi = \pm \pi $ the maximum is slightly enhanced.

The answer as to why the $x$ component of the electric field does not contribute to the magnonic spin current is inferred from  the magnon dispersion relations
\begin{equation}
\displaystyle \omega = \gamma H_{0} + \frac{2 \gamma A_{ex}}{\mu_{0} M_{s}}k_{\pm}^{2} \pm\frac{2 \gamma c_{E}}{\mu_{0} M_{s}} (E_{y} \text{sin}\theta\text{sin}\phi -E_{z} \text{sin}\theta\text{cos}\phi) k_{\pm}.
\label{eq_S1}
\end{equation}
As evident,  $E_{x}$ does not appear in the dispersion relations and hence it is irrelevant  when it comes to the asymmetry of the left and right propagating magnons.

\section{Thermal gradient vs. $E$-field gradient}
\label{part7}

\begin{figure}
	\centering
	\includegraphics[width=0.49\textwidth]{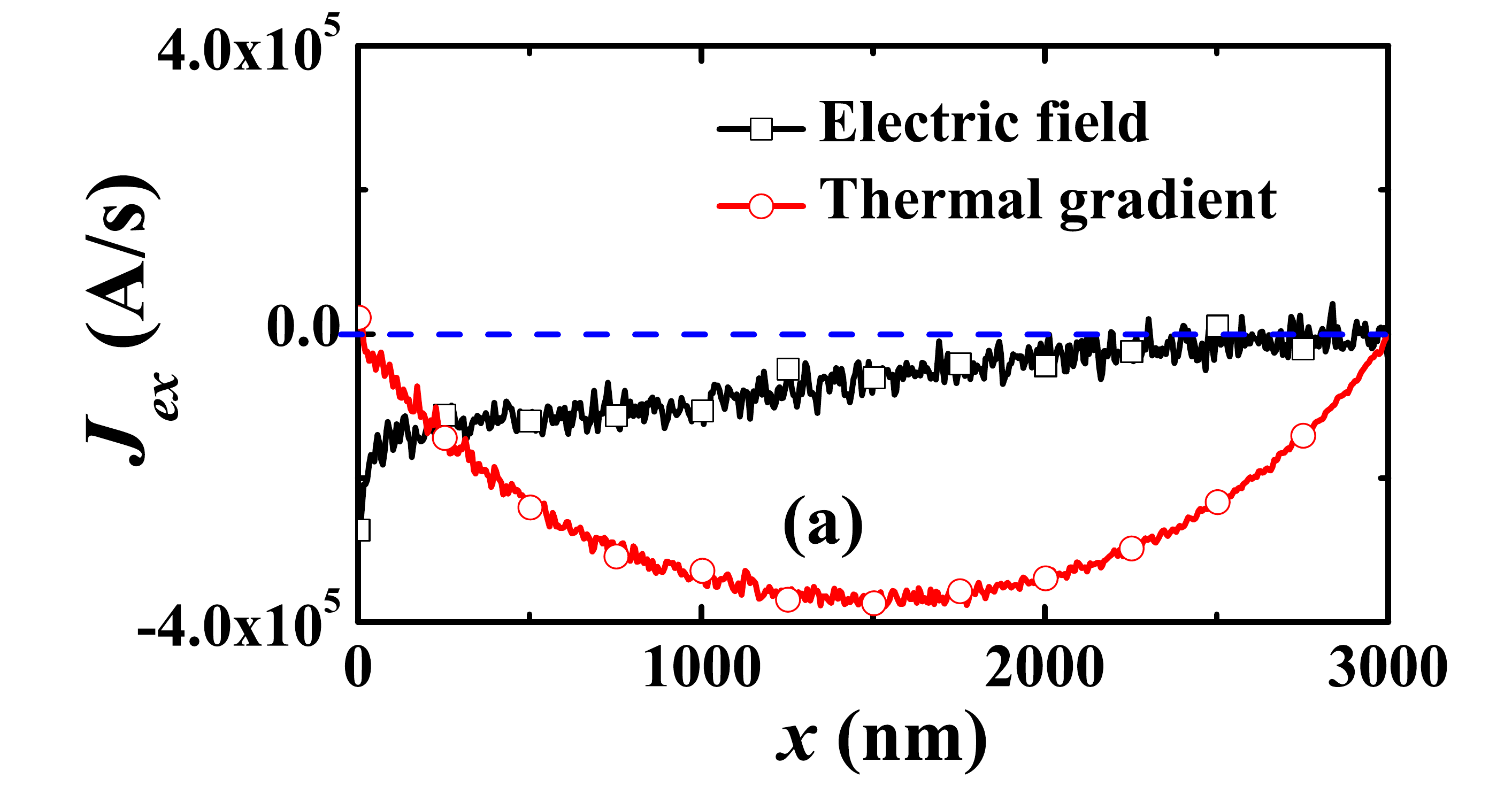}
	\includegraphics[width=0.49\textwidth]{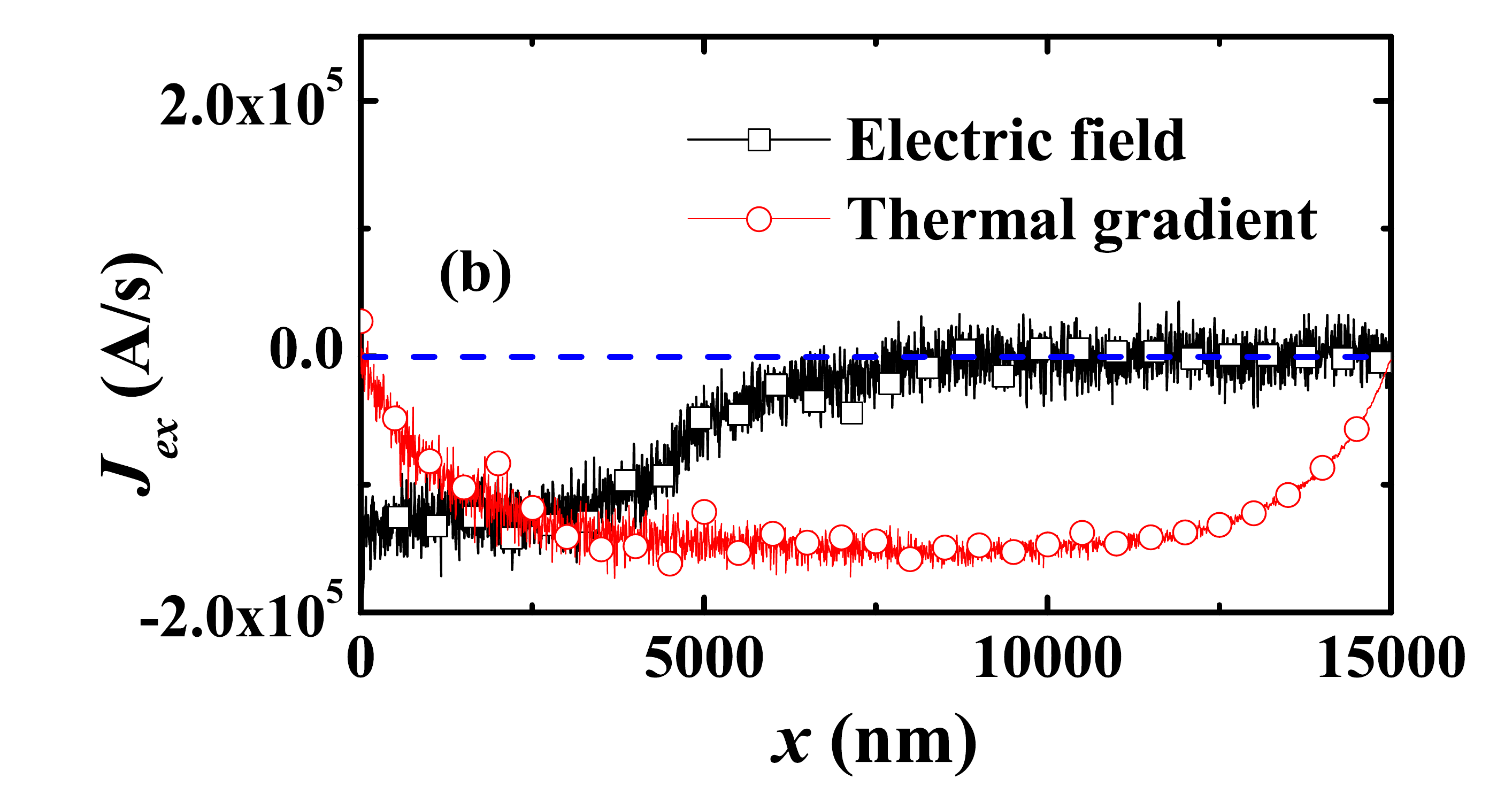}	
	\caption{(a) Profile of the exchange spin current $ J_{ex} $ generated by a temperature gradient and an electric field ($ E_{y}=-0.05 $ MV/cm in the left subsystem $ 0 \leq x \leq 900 $ nm). The length of sample is 3000 nm. (b) Profile of the exchange spin current $ J_{ex} $ generated by a temperature gradient and an electric field ($ E_{y}=-0.05 $ MV/cm in the left subsystem $ 0 \leq x \leq 4500 $ nm). The length of sample is 15000 nm.}
	\label{fig_S4}
\end{figure}

Above we uncovered  the mechanism for the $E$-field-gradient-induced magnonic spin current at finite $T$. Such a current can also be achieved without $E$ fields but by applying a temperature gradient.
Sizable effects  entail  inducing  large $T-$gradients on the nanoscale, which is evidently more challenging than just applying  a uniform $T$ and a static $E$-field. In addition, $E$ field gradient might be intrinsically present for instance in heterostructures. Extensive numerical calculations show that the achieved spin current in our case is substantial making the present study a competitive  alternative to applying $T-$gradients. A comparison between  generating  spin currents via $E$-field or $T$ gradients is shown in Fig. \ref{fig_S4}(a), (b). We numerically calculated the spin current generated by applying a temperature gradient $ T(x) = 15 - \frac{15 x}{3000 \mathrm{nm}}$ K (we used the standard recipe for treating the magnonic spin Seebeck effect \cite{Etes}), and a spin current generated by applying nonuniform electric field and uniform temperature of the same order $T=15$K. As we see, the amplitudes of the spin currents are quantitatively match each others.
In fact, both the present   and the standard approach to the magnonic spin Seebeck effect share some fundamental aspects: Energetically, the magnonic excitations and their decay are associated with an energy exchange with the (phononic  bath).  The same applies to the  angular momentum flow associated with the spin current. Both aspects enter our model through the Gilbert damping. \textbf{For a large scale area (15000 nm in our case) the magnonic spin current (induced by $E$ field gradient ) can not spread in the whole sample due to the magnon attenuation. For an experimental realization with large scale system one may increase the number of the interfaces \cite{wang2017} (i. e., excitation sources). With several excitation sources  the electric field induced magnonic excitation mechanism is still operational. }

\section{Conclusions }
\label{part8}
 
We uncovered a new method for converting thermal fluctuations in ferromagnetic insulators to a directed spin current and an emergent \textit{effective} electric polarization by means of a static, electric field gradient.  The key point is the local modulation  in the magnons dispersions and the magnon density profiles that are brought about by the $E$ field gradient.
On the scale of this gradient one may use the local magnon density to define a local magnon temperature  which is different from that related to the external bath, resulting so in a gradient of the magnon temperature. Viewed from this angle, the emergence of the spin current comes as no surprise, and it is also comprehensible  that its magnitude is comparable to that of the spin Seebeck current generated by an external temperature gradient. The change in the magnon dispersions by the electric field is also important, as magnons are  discriminated  according to their wave vector, giving the current so its direction. We presented analytical as well as full numerical results testifying  the predicted phenomena and evidencing  that   the magnonic spin current is substantial and controllable  by a moderate external electric field, pointing so to new avenues for applications.

\section{Acknowledgments:} This work was supported by the German Science Foundation, DFG under SFB 762 as well as by the National Natural Science Foundation of China under Grants No.11704415, No. 11674400 and No. 11374373.

\appendix

\section{Material parameters}

The material parameters used in numerical calculations are adopted for iron garnets $ (\mathrm{BiR})_3(\mathrm{FeGa})_5\mathrm{O}_{12} $ (R = Lu, Tm) \cite{ASLo, VRis}: $M_S = 1 \cdot 10^4$ A/m, $A_{ex} = 5 \cdot 10^{-12}$ J/m, and Gilbert damping constant $ \alpha = 0.005 $.
In order to assess and remove possible numerical spurious effects, such as the choice of the initial state and boundary conditions, we start with a randomly chosen magnetic state. However, we propagate the magnetization within the time interval that exceeds the relaxation time of the homogeneously magnetized FM. Results obtained by us corresponds to the steady state equilibrated magnetization dynamics and therefore spin current does not depend on time.
An external magnetic field $ \vec{H}_{\text{ext}} = H_{0} (\text{cos}\theta,\text{sin}\theta\text{cos}\phi,\text{sin}\theta\text{sin}\phi) $, is of the order of $ H_{0} = 6 \times 10^{5} $ A/m and induces uniform ground state magnetization $ \vec{M}_{0} = M_{s} (\text{cos}\theta,\text{sin}\theta\text{cos}\phi,\text{sin}\theta\text{sin}\phi) $ with the polar $ \theta = \pi / 2 $ and the azimuthal $ \phi $ angles. Compared to the external field influence, the effect of the material anisotropy is small enough and can be neglected (the effective anisotropy field estimated for the anisotropy constant $ K = 500 $ J/m$ ^3 $ is of the order of $ 8 \times 10^{4} $ A/m ). The coupling constant $ c_{E} $ determines the amplitude of the saturated polarization $ \vec{P} $. The saturated polarization calculated via the micromagnetic simulation is of the order of $ 1000 \mu $ C/m$ ^2 $ for $ c_{E} = 6 $ pC/m. The value of the parameter $ c_E $ is adopted from the experimental data \cite{ASLo}. In spin-driven helical multiferroics (for example, orthorhombic manganites RMnO$ _3 $) the polarization induced by the non-collinear magnetic texture is typically  of the order of $ 1000 \mu $ C/m$ ^2 $ \cite{TAoy}. The amplitude of the external electric field considered in the micromagnetic simulations is varied within the interval $ (-0.05, 0.05) $  MV/cm. The corresponding effective DMI constant $ D_E $ is of the order of $ (-0.03, 0.03) $  mJ/m$ ^2 $. At  zero temperature $ T = 0 $ K, the  magnetic order in the system is ferromagnetic and $\vec{P}=0$.

\end{document}